\documentclass[12pt]{article}
\pdfoutput=1
\usepackage{epsf,amsfonts,amssymb,epsfig,amsmath,mathtools,graphics,slashed}
\usepackage{hyperref,hep,graphicx,subfig}
\usepackage{rotating}

\newcommand{\nn}{\notag \\}

\newcommand{\s}{\sigma}

\makeatletter

\@addtoreset{equation}{section}
\makeatother

\begin{document}

\begin{titlepage}

\vfill

\begin{flushright}
Imperial/TP/2017/JG/01\\
DCPT-16/37
\end{flushright}

\vfill

\begin{center}
   \baselineskip=16pt
   {\Large\bf DC Conductivity and Higher Derivative Gravity }
  \vskip 1.5cm
  \vskip 1.5cm
Aristomenis Donos$^1$, Jerome P. Gauntlett$^2$\\ Tom Griffin$^2$ and Luis Melgar$^2$\\
     \vskip .6cm
     \begin{small}
      \textit{$^1$Centre for Particle Theory and Department of Mathematical Sciences\\Durham University, Durham, DH1 3LE, U.K.}
        \end{small}\\    
         \begin{small}\vskip .6cm
      \textit{$^2$Blackett Laboratory, 
  Imperial College\\ London, SW7 2AZ, U.K.}
        \end{small}\\
        \end{center}
     \vskip .6cm
\begin{center}
{\it Dedicated to the 75th birthday of John H. Schwarz}
\end{center}
\vfill

\begin{center}
\textbf{Abstract}
\end{center}
\begin{quote}
For Gauss-Bonnet gravity and in the context of holography we show how the thermal DC conductivity can be obtained
by solving a generalised system of Stokes equations for an auxiliary fluid on a curved black hole horizon.
For more general higher derivative theories of gravity coupled to gauge-fields, we also analyse the linearised thermal and electric currents that are
produced by DC thermal and electric sources. We show how suitably defined DC transport current fluxes of the dual field theory
are given by current fluxes defined at the black horizon. 

\end{quote}

\vfill

\end{titlepage}

\setcounter{equation}{0}
\section{Introduction}

The thermal and electric conductivities are important observables to study within the framework of holography.
For Einstein-Maxwell theory, possibly coupled to other matter fields, it has been shown  
that the thermoelectric DC conductivity of the dual field theory at finite temperature can be obtained by solving a 
system of Stokes equations for an auxiliary fluid on a black hole horizon \cite{Donos:2015gia,Banks:2015wha,Donos:2015bxe}. 
The purpose of this paper is to
investigate how this striking result can be extended to theories of gravity whose Lagrangians contain higher derivative terms.
Such higher-derivative terms naturally arise in string and M-theory, leading to finite coupling and $1/N$ effects in the context of holography,
and our results provide useful tools to investigate their effect on the conductivity, analogous to similar investigations for the shear viscosity \cite{Brigante:2007nu}.

We begin by summarising some key aspects of the analysis in \cite{Donos:2015gia,Banks:2015wha,Donos:2015bxe}.
We first recall that the natural framework for studying DC conductivities in holography is provided by
holographic lattices \cite{Hartnoll:2012rj,Horowitz:2012ky,Donos:2012js,Chesler:2013qla,Donos:2013eha,Andrade:2013gsa}. 
These are stationary black hole spacetimes with Killing horizons that are dual to CFTs in thermal equilibrium which have been
deformed by operators that explicitly break the translation invariance of the CFT. Although not essential, one often considers spacetimes with a single black hole horizon of planar topology and the deformations are taken to be periodic in the spatial directions.
The explicit breaking of the translation symmetry is imposed by demanding suitable boundary conditions on the bulk fields at the AdS boundary;
this is essential, generically, in order to obtain a finite DC response. 

To calculate the thermoelectric DC conductivity one analyses a linear perturbation of the bulk fields about the background holographic lattice.
The boundary conditions that are imposed on the perturbation at the AdS boundary are associated
with the application of external DC thermal and electric sources to the dual field theory. An important result of \cite{Donos:2015bxe}, building on \cite{PhysRevB.55.2344,Blake:2015ina,Hartnoll:2007ih},
was to correctly identify the conserved transport currents of the dual field theory. These are obtained from the usual holographic currents
by suitably subtracting off terms that arise when the equilibrium dual field theory, described by
the background holographic lattice, has non-vanishing magnetisation currents.
Furthermore, it was also shown in \cite{Donos:2015gia,Banks:2015wha,Donos:2015bxe} how the transport currents can be obtained from the perturbed bulk geometry, including a contribution
from currents defined at the black hole horizon. Importantly, the currents at the horizon, which are conserved, only involve a subset of the bulk perturbation and they take the form of constitutive relations for an auxiliary fluid. It was also shown in \cite{Donos:2015gia,Banks:2015wha,Donos:2015bxe}
that the current fluxes at the horizon, which in the case of periodic lattices are just the zero
modes of the currents, are equal to the transport current fluxes of the dual field theory. This last result implies that if one knows
the currents at the black hole horizon as functions of the applied DC sources on the horizon\footnote{For simplicity, here we are assuming
that the holographic lattice has a globally defined radial coordinate outside the black hole horizon and in this case we can take the applied DC
sources to be independent of the radial coordinate. For a more general discussion see \cite{Banks:2015wha,Donos:2015bxe}.} then
one can obtain the transport current fluxes of the dual field theory as functions of the DC sources and hence, by definition, the DC conductivity.

In sections \ref{gendisc}-\ref{gbonsec} we generalise all of these results to general theories of gravity coupled to a gauge field; the extension to include additional matter fields is straightforward. We will focus on holographic lattices, but some of our results are independent of the asymptotic boundary conditions 
and may have applications outside of holography.
In section 2, by carrying out a Kaluza-Klein reduction of the gravity theory on the time direction\footnote{To do this 
we need to introduce the DC sources in a slightly different gauge than the ``linear in time" perturbations that were used in \cite{Donos:2015gia,Banks:2015wha,Donos:2015bxe} (building on \cite{Donos:2014uba,Donos:2014cya}).} we can obtain natural definitions of the electric and thermal transport currents of the dual field theory with the properties mentioned in the previous paragraph. 
In particular, we present a recipe for obtaining the conserved currents at the horizon as well as showing that the current fluxes
at the horizon are the same as the transport current fluxes of the dual field theory. 

To obtain explicit expressions for the currents at the horizon in terms of the perturbation requires more information
about the specific theory of gravity that is being considered.
In section \ref{emcasesec}, for the case of Einstein-Maxwell theory in $D$ spacetime dimensions and stationary black holes, we first show that we recover exactly the same currents on the horizon that were derived in \cite{Donos:2015bxe}. In section \ref{gbonsec} we consider the case of pure gravity in $D$ spacetime dimensions including curvature squared contributions. A special case of this class of theories includes Gauss-Bonnet 
gravity in $D\ge 5$, which we return to below. For the
class of static black holes, for which there are no background 
magnetisation currents, we derive explicit expressions for the conserved heat current at the horizon.

Another key aspect of \cite{Donos:2015gia,Banks:2015wha,Donos:2015bxe}, for theories of gravity without higher derivative terms, was the use of a radial decomposition of the bulk equations 
of motion to evaluate the momentum, Hamiltonian and Gauss-law constraints on the horizon\footnote{More precisely, the constraints
are evaluated on a hypersurface of fixed radial coordinate just outside the horizon, a ``stretched horizon", and then a limit is taken.}
for the DC perturbed spacetime. It was shown that this leads to a set of linearised, time-independent, forced Navier-Stokes equations for a charged, incompressible fluid on the black hole horizon. These equations, which we refer to as Stokes equations, also include the
conservation laws for the electric and heat currents on the horizon, that we already discussed above.
A crucial feature of the system of Stokes equations is that they form a closed set of equations for a subset of the perturbation on the horizon. 
By solving the Stokes equations one can obtain the conserved currents at the horizon in
terms of the DC sources and hence the DC conductivity of the dual field theory. 
The way in which the Stokes equations appear on the horizon is related to the old membrane paradigm of \cite{Price:1986yy}.
It is worth noting, though, that for the linearised perturbations driven by a DC source studied in \cite{Donos:2015gia,Banks:2015wha,Donos:2015bxe}
one obtains linearised Stokes equations for 
an incompressible fluid on a curved horizon, whereas in \cite{Price:1986yy} there are equations for a compressible fluid with negative bulk viscosity.

The approach of \cite{Donos:2015gia,Banks:2015wha,Donos:2015bxe} for 
obtaining the Stokes equations at the horizon can immediately be deployed to study  
theories of gravity with higher derivative terms that have a well-defined Cauchy problem for radial evolution.
The Lovelock theories of pure gravity, which have equations of motion only involving second order time derivatives of the metric,
satisfy this condition. We will focus on the particular case of Gauss-Bonnet theory in $D\ge 5$ spacetime dimensions, 
whose Lagrangian consists of the usual Einstein-Hilbert term and a negative cosmological constant, supplemented by
the Gauss-Bonnet term with coupling $\tilde\alpha$ as given in \eqref{hdact2}. 
In section \ref{gbns} we will show that when the constraints are evaluated on the horizon we obtain
a generalised set of Stokes equations which have non-trivial dependence on $\tilde\alpha$. 
Taking a perturbative approach\footnote{The perturbative approach is directly relevant to applications to string theory. It is also worth
noting that attempting to consider the Gauss-Bonnet terms non-perturbatively leads to issues with causality \cite{Camanho:2014apa}. } to the higher derivative Gauss-Bonnet theory, it is of particular interest to obtain the leading
order corrections in $\tilde\alpha$. If $h_{ij}$ is the horizon metric,  
with Levi-Civita connection $\nabla_i$, the leading order Stokes equations can be written\footnote{We use a slightly different notation in the main text.}
\begin{align}\label{icintro}
\nabla_i(\delta^i_j-4\tilde\alpha {G}^i_{j})v^j&=0\,,\nn
-2\nabla^i\left(S_{ij}^{kl}\nabla_k v_l\right)&=(\delta^i_j-4\tilde\alpha {G}^i_{j})\left(4\pi T\zeta_i-{\nabla_i p}\right)\,,
\end{align}
where $G_{ij}=R_{ij}-R h_{ij}$ is the Einstein tensor on the horizon and 
\begin{align}\label{essintro}
S^{kl}_{ij}=\left[1-\tilde\alpha 2(D-4)(D-1)\right]\delta^{(k}_i\delta^{l)}_j
-\tilde\alpha\left[2h^{}_{ij}R_{}^{kl}+4\delta^{(k}_{(i}R_{}{}^{l)}_{j)}+4R_{}{}_i{}^{(k}{}_j{}^{l)}\right]\,.
\end{align}
Note that the horizon metric quantities depend on $\tilde \alpha$ and need to be expanded in these expressions.

It is interesting to highlight that the coefficient appearing in the first term of \eqref{essintro} is precisely the same as the 
the ratio of the shear viscosity to entropy ratio that was calculated for a planar black brane in
Gauss-Bonnet theory in \cite{Brigante:2007nu}, namely $4\pi\frac{\eta}{s}=1-\tilde \alpha 2(D-4)(D-1)$.
The reason for this will be explained in section \ref{gbns}. 
As in \cite{Donos:2015gia,Banks:2015wha,Donos:2015bxe}, we emphasise that
\eqref{icintro},\eqref{essintro}
again form a closed set of equations for a subset of the perturbation on the horizon and allow one to
obtain the conserved currents at the horizon in terms of the DC sources and hence, in turn, the DC conductivity of the dual field theory.
Thus the membrane paradigm for DC response of \cite{Donos:2015gia,Banks:2015wha,Donos:2015bxe} generalises to Gauss-Bonnet 
gravity\footnote{We note that the membrane paradigm of \cite{Price:1986yy} was adapted to Gauss-Bonnet gravity 
in \cite{Jacobson:2011dz}.}.
We will also show in section \ref{gbns} how the new Stokes equations can be solved
in the special case that there is only dependence on one of the spatial directions on the horizon.

We briefly conclude in section \ref{fincom}
and we have several appendices containing various technical material.

\section{Bulk physics in the DC limit}\label{gendisc}

\subsection{General set-up}
We consider a general bulk theory in $D$ spacetime dimensions that 
is diffeomorphism invariant and has a local internal $U(1)$ gauge symmetry. 
More precisely, we will assume that the Lagrangian, ${\cal L}$, is gauge-invariant and transforms as a scalar under diffeomorphims.
We further assume that the theory only depends on a metric and an abelian
gauge field, but we note that it is straightforward
to include additional matter fields and other gauge symmetries. 
We thus consider
\begin{align}
S=\int d^Dy\sqrt{-g}\mathcal{L}(g_{\mu\nu},A_\mu)\,,
\end{align}
and we note that for notational simplicity we have suppressed the dependence of $\mathcal{L}$ on the derivatives of the fields.
In such a theory, we are interested in perturbing stationary charged black hole backgrounds by DC
electric fields and thermal gradients and examining the currents that are produced.

We begin by writing the $D$-dimensional coordinates as 
\begin{align}
y^\mu=(t,y^m)\,,
\end{align}
where the coordinates $y^m$
parametrise a $(D-1)$-dimensional manifold $M_{D-1}$. We will assume that $\partial_t$ is a Killing vector that leaves the gauge field invariant and hence consider the general ansatz:
\begin{align}\label{eq:diff_metric_gfield_ansatz}
ds^{2}&=-H^{2}\,\left( dt+\alpha \right)^{2}+ds^{2}\left(M_{D-1} \right)\,,\nn
A&=A_{t}\,\left( dt+\alpha \right)+\beta\,.
\end{align}
Here $ds^{2}\left(M_{D-1}\right)\equiv \gamma_{mn}dy^mdy^n$, while $\alpha=\alpha_m dy^m$, $\beta=\beta_m dy^m$ 
are one-forms and $H$ is a function, all defined on $M_{D-1}$. This ansatz is sufficient to accommodate both the stationary black hole backgrounds that we are interested in as well as the DC perturbation we wish to consider. 

We will keep the discussion general for the moment, making no further assumptions on the background. 
However, at some point we will consider a $D$-dimensional background black hole spacetime, with a Killing horizon, and then the coordinates $y^m=(r,x^i)$ will parametrise a holographic radial direction, $r$, as well as the spatial directions, $x^i$, of the dual field theory. Thus $M_{D-1}$
will have a holographic boundary. In addition, in order to consider $M_{D-1}$ as a manifold, with non-trivial boundary, we can
envisage the radial direction to be terminated at a ``stretched horizon", located at a very small distance from the event horizon. When
we discuss global issues concerning regularity of the spacetime at the horizon, we will, of course, discuss them in the context of the full $D$-dimensional spacetime.

We now dimensionally reduce our theory in $D$ spacetime dimensions on the time direction. The equations of motion for the fields on
$M_{D-1}$ can be obtained from a $D-1$ dimensional action of the form
\begin{align}\label{eq:lower_D_functional}
S=\int_{M_{D-1}}\,d^{D-1}y\,\sqrt{\gamma_{D-1}}\,H\,\mathcal{L}\left(H,A_{t},\alpha,\beta,\gamma_{mn} \right)\,,
\end{align}
where $\cal L$ is a local function of the fields appearing in the ansatz \eqref{eq:diff_metric_gfield_ansatz}, as well as the spatial derivatives of the fields. For notational simplicity we have again suppressed the dependence of $\mathcal{L}$ on
the spatial derivatives of the fields.

There are several restrictions on $\mathcal{L}$ which follow
from diffeomorphism and gauge invariance of the $D$-dimensional theory.
Firstly, local coordinate transformations of the form $t\rightarrow t+\Lambda_{E}(y^m)$ imply that $\mathcal{L}$ is invariant under 
\begin{align}
\alpha\rightarrow \alpha+d\Lambda_{E}(y^m)\,.
\end{align}
Second, the gauge transformations $A(t,y^m)\rightarrow A(t,y^m)+d\Lambda_{M}(t,y^m)$ with $\Lambda(t,y^m)=c\,t+\lambda_{M}(y^m)$, for constant $c$, imply that
$\mathcal{L}$ is invariant under
\begin{align}
A_{t}\rightarrow A_{t}+c\,,\qquad
\beta\rightarrow \beta+d\lambda_M-c\alpha\,.
\end{align}
Third, invariance under time scalings, $t\rightarrow \lambda\,t$  imply that $\mathcal{L}$ is invariant under
\begin{align}
H\rightarrow \lambda\,H,\qquad \alpha\rightarrow\alpha/\lambda,\qquad A_{t}\rightarrow \lambda\,A_{t}\,.
\end{align}

These conditions imply that $\cal L$ will have the following functional dependence:
\begin{align}\label{ellform}
&\mathcal{L}\left(H,A_{t},\alpha,\beta,\gamma_{mn} \right)=\mathcal{L}\left(h,u,v,w,\gamma_{mn}\right)\,,
\end{align}
where $h,u$ are one-forms and $v,w$ are two-forms, all defined on $M_{D-1}$, defined by
\begin{align}\label{eq:forms_def}
&h=d\ln H,\qquad
u=H^{-1}dA_{t}, \qquad
v=d\beta+A_{t}\,d\alpha,\qquad
w=H\,d\alpha\,.
\end{align}
Another restriction is that $\cal L$ must be invariant under $(u,w)\to -(u,w)$. This follows 
from the fact that the dimensional reduction ansatz is invariant
under $H\to -H$.

The equations of motion for $\alpha$ and $\beta$ will play an important role in the sequel. For simplicity, for now we will assume that $\cal L$ does not depend on derivatives of the one- and two-forms. This is the situation that arises for
theories of gravity with two derivatives, such as Einstein-Maxwell theory. The simple generalisation that arises for higher derivative theories of gravity will be presented later in section \ref{gbonsec}. With this assumption we find that
the equations of motion for $\alpha$ and $\beta$ are given by
\begin{align}\label{formeqmotion}
&\nabla_{m} V^{mn} =0,\qquad \nabla_{m} W^{mn} =0\,,
\end{align}
where 
\begin{align}\label{eq:form_eoms}
V^{mn}=H\,\frac{\delta{\mathcal{L}}}{\delta v_{mn}},\qquad
W^{mn}=H^{2}\,\frac{\delta{\mathcal{L}}}{\delta w_{mn}}+HA_{t}\frac{\delta{\mathcal{L}}}{\delta v_{mn}}
\,.
\end{align}
In other words the two-forms $W=\tfrac{1}{2}W_{mn}dx^m\wedge dx^n$ and $V=\tfrac{1}{2}V_{mn}dx^m\wedge dx^n$, with indices lowered with $\gamma_{mn}$, are both co-closed. This is a key result.
As we will see, $V$ and $W$ are associated with the electric and heat currents of the dual field theory in the context
of holography.

\subsection{DC perturbation}
We now want to consider the equations that govern a specific linear perturbation about 
a given stationary background solution to the equations of motion. We write the stationary background
solution as
\begin{align}
&H=H^{(B)}, \quad \alpha=\alpha^{(B)},\quad \gamma_{mn}=\gamma^{(B)}_{mn}\,,\quad
A_{t}=A_{t}^{(B)},\quad \beta=\beta^{(B)}\,,
\end{align}
with all background quantities independent of the time coordinate $t$.

The perturbation we want to consider is seeded by two closed one-forms
$\zeta,E$ that are globally defined on $M_{D-1}$. To achieve this 
we will introduce two locally defined functions, 
$\phi_{T}$ and $\phi_{E}$, on $M_{D-1}$ via\footnote{Note that
$\phi_{T}, \phi_{E}$ can depend, in general, on all spatial coordinates. Later, in section \ref{secbhgih}, 
to simplify the presentation, we will take $\phi_{T}$ and $\phi_{E}$ to be independent of the radial coordinate.}
\begin{align}
\zeta=d\phi_{T}\,,\qquad E=d\phi_{E}\,.
\end{align}
The perturbation that we consider is then given by
\begin{align}\label{expsII}
&H=H^{(B)}\left(1-\phi_{T}\right) +\delta H, \quad \alpha=\alpha^{(B)} \left(1+\phi_{T} \right)+\delta \alpha\,,
\quad \gamma_{mn}=\gamma_{mn}^{(B)}+\delta \gamma_{mn}\,,\notag\\
&A_{t}=A_{t}^{(B)}\,\left(1-\phi_{T} \right)+\phi_{E}+\delta A_{t},\qquad
\beta=\beta^{(B)}-\phi_{E}\,\alpha^{(B)}+\delta\beta\,,
\end{align}
with $\delta H,\delta\alpha,\delta A_t,\delta\beta$ and $\delta \gamma_{mn}$ all globally defined perturbations on $M_{D-1}$. Thus, at linearised order, the perturbed metric and gauge-field takes the form
\begin{align}
ds^2+\delta(ds^2)&=-H^{(B)2}(1+2H^{(B)-1}\delta H)[(1-\phi_T)dt+\alpha^{(B)}+\delta\alpha]^2
+(\gamma^{(B)}_{mn}+\delta\gamma_{mn})dx^mdx^n\,,\nn
A+\delta A&=(A^{(B)}_t+\delta A_t)[(1-\phi_T)dt+\alpha^{(B)}+\delta \alpha]+\beta^{(B)}+\delta\beta+\phi_E dt\,.
\end{align}

For the holographic applications we consider later, on the holographic boundary the one-form $E$ parametrises an applied DC electric field
source term, while $\zeta$ parametrises a DC thermal gradient via\footnote{The sign can be established as follows. Consider the metric $ds^2=-(1-2\phi_T)dt^2+dx^idx^i$. For $\dot x^i<<1$ and perturbative in $\phi_T$, the geodesic equation gives $\ddot x^i=\zeta_i$. Since heat moves from hot to cold we identify
 $\zeta$ with $-T^{-1}dT$.}: $\zeta\leftrightarrow -T^{-1}dT$, where $T$ 
is the locally defined temperature.
After a simple
coordinate transformation and gauge transformation (see \eqref{coordandgauge} below) we can easily show that this gives rise
to the perturbation containing terms linear in the time coordinate $t$ that have been used in derivations of the DC 
conductivity, starting with \cite{Donos:2014uba,Donos:2014cya}. As we will now see, it is convenient to work with the above locally defined, time independent perturbation so that we can directly utilise the Kaluza-Klein dimensional reduction formulae.
Importantly, the key equations we are ultimately interested in will only involve the globally defined one-forms $(\zeta,E)$ and not the locally defined 
functions $(\phi_{T},\phi_{E})$. We will make some additional comments on this in the next sub-section.

To proceed we now write out the perturbation in terms of the one- and two-forms defined in \eqref{eq:forms_def}. We find:
\begin{align}\label{eq:forms_pert}
\delta h&=d(H^{(BG)-1}\delta H)-\zeta\,,\nn
\delta u&=H^{(B)}{}^{-1}d\delta A_{t}-H^{(B)}{}^{-2}\,\delta H \,d A_{t}^{(B)}-H^{(B)}{}^{-1}\,A_{t}^{(B)}\,\zeta+H^{(B)}{}^{-1}\,E\,,\nn
\delta v&=d\delta\beta+\delta A_{t}\,d\alpha^{(B)}+A_{t}^{(B)}\,d\delta\alpha+A_{t}^{(B)}\,\zeta\wedge\alpha^{(B)}-E\wedge\alpha^{(B)}\,,\nn
\delta w&=\delta H\,d\alpha^{(B)}+H^{(B)}\,d\delta\alpha+H^{(B)}\,\zeta\wedge \alpha^{(B)}\,,
\end{align}
and, in particular, we observe that they involve the globally defined one-forms $(\zeta,E)$, as just mentioned.

We next examine how the two-forms $V$ and $W$ defined in \eqref{eq:form_eoms} depend
on the linearised perturbation.
Starting with $V$ we find
\begin{align}\label{delvee}
\delta V^{mn}=&-V^{(B)mn}\,\phi_{T}+\delta L^{mn}\,,
\end{align}
where we have defined 
\begin{align}\label{eq:L_def}
\delta L^{mn}
\equiv \left(\frac{\delta{\mathcal{L}}}{\delta v_{mn}}\right)^{(B)}\,\delta H+H^{(B)}\,\delta\left(\frac{\delta{\mathcal{L}}}{\delta v_{mn}}\right)\,.
\end{align}
The two-form $\delta L_{mn}$ is globally defined on $M_{D-1}$.
This is because $\mathcal{L}$ is a functional of the one- and two-forms in \eqref{eq:forms_def} which we have shown, in equation \eqref{eq:forms_pert}, remain well defined after perturbing under \eqref{expsII}.  While $\phi_{T}$ does appear in $\delta V^{mn}$, it drops out of the equations of motion
\eqref{formeqmotion}. Specifically, after using the background equation of motion,
$\partial_{m}[(\gamma_{D-1}^{(B)})^{1/2}\, V^{(B)mn}]=0$, we find that \eqref{formeqmotion} implies
\begin{align}\label{emcase}
\partial_{m}\left((\gamma_{D-1}^{(B)})^{1/2}\, \delta L^{mn}+\delta[(\gamma_{D-1})^{1/2}]\, V^{(B)mn}\right)=(\gamma_{D-1}^{(B)})^{1/2}\zeta_{m} V^{(B)mn}\,.
\end{align}

Following a similar logic for the two-form $W$ we find
\begin{align}\label{deldub}
\delta W^{mn}=&-2\,W^{(B)mn}\,\phi_{T}+V^{(B)mn}\,\phi_{E}+\delta K^{mn}\,,
\end{align}
where we have defined
\begin{align}\label{eq:K_def2}
\delta K^{mn}\equiv &\,2H^{(B)}\,\left(\frac{\delta{\mathcal{L}}}{\delta w_{mn}}\right)^{(B)}\delta H+H^{(B)}{}^{2}\,\delta \left(\frac{\delta{\mathcal{L}}}{\delta w_{mn}}\right)\nn
&+H^{(B)}\left(\frac{\delta{\mathcal{L}}}{\delta v_{mn}}\right)^{(B)}\,\delta A_{t}
+A^{(B)}_{t}\delta L^{mn}\,.
\end{align}
The two-form $\delta K_{mn}$ is again globally defined on $M_{D-1}$. 
While $(\phi_{T},\phi_{E})$ do appear in $\delta W^{mn}$ they again also drop out of the equations of motion
\eqref{eq:form_eoms}. Specifically, after using the background equation of motion,
$\partial_{m}[(\gamma_{D-1}^{(B)})^{1/2}\, W^{(B)mn}]=0$, we find that \eqref{eq:form_eoms} implies
\begin{align}\label{eq:GR_div_eqn}
&\partial_{m}\left((\gamma_{D-1}^{(B)})^{1/2}\, \delta K^{mn}+\delta[(\gamma_{D-1})^{1/2}]\, W^{(B)mn}\right)=
(\gamma_{D-1}^{(B)})^{1/2}\left[2\,\zeta_{m}W^{(B)mn}-\,E_{m}V^{(B)mn}\right]\,.
\end{align}

The two equations \eqref{emcase} and \eqref{eq:GR_div_eqn} are key results. As we will see below, they generalise the equations given in eq. (5.6) of \cite{Donos:2015bxe} from a specific theory of gravity to
a more general one and, in addition, it is clear how to extend the theory to include additional matter fields.

\subsection{Black hole geometry in holography}\label{secbhgih}
We now consider the results in the previous subsection in a holographic context of DC perturbations about
a background black hole geometry.
While it is straightforward to be significantly more general, as in section 5 of \cite{Donos:2015bxe}, to simplify the presentation
we now assume that the stationary background solution has a single black hole Killing horizon and we will also assume that the solution has a globally defined radial coordinate, $r$, outside the black hole. Introducing the coordinates
\begin{align}
y^m=(r,x^i)\,,
\end{align}
we assume that the black hole is located at $r=0$ and that there is also a boundary
at $r=\infty$, which we take to approach a suitably deformed $AdS_D$. More precisely, as $r\to \infty$ we have
\begin{align}\label{asmet}
ds^2&\to r^{-2}dr^2+r^2\left[g^{(\infty)}_{tt}dt^2+g^{(\infty)}_{ij}dx^idx^j+2g_{ti}^{(\infty)}dtdx^i\right]\,,\nn
A&\to A^{(\infty)}_t dt+A^{(\infty)}_i dx^i\,, 
\end{align}
where $g^{(\infty)}_{tt}$ etc. are functions of the spatial coordinates, $x^i$, only, 
and parametrise the spatially dependent sources for the stress tensor and abelian current operators that have been
used to deform the CFT. Such black hole solutions are known
as holographic lattices. Often we are interested in CFTs on spacetimes with planar topology
with deformations that are periodic in the spatial directions, so-called ``periodic holographic lattices". 
In the set-up of this paper, by considering a fundamental domain, this corresponds, effectively, to considering CFTs living on a $(D-2)$-dimensional torus with a non-trivial metric, associated with a source for the stress tensor, as
well as additional periodic sources for the abelian current, the simplest of which is a periodic chemical potential $A^{(\infty)}_t$.

The manifold $M_{D-1}$ is defined for $0<r<\infty$ and the associated metric can be 
written
\begin{align}\label{eq:base_metric}
ds^{2}(M_{D-1})=\gamma_{mn}^{(B)}(y^p)dy^m dy^n\,.
\end{align}
This metric becomes singular at the location of the black hole at $r=0$. 
In order that the $D$-dimensional black hole solution is regular at the horizon, which is Killing with respect to $\partial_t$, 
we demand that as $r\to 0$:
\begin{align}\label{nhorexp}
H^{(B)}{}^{2}= 4\pi T\,r+\mathcal{O}(r^{2})\,,\quad
\gamma_{rr}^{(B)}=\frac{1}{4\pi T\,r}+\mathcal{O}(1)\,,\quad
A_{t}^{(B)}= \,a_{t}^{(0)}(x)\,r+\mathcal{O}(r^{2})\,,
\end{align}
with the remaining quantities given by
\begin{align}\label{nhorexp2}
\alpha^{(B)}_{r}&=\alpha^{(0)}_r+\mathcal{O}(r)\,,\qquad
\alpha^{(B)}_{i}=\alpha^{(0)}_i+\mathcal{O}(r)\,,\notag\\
\gamma^{(B)}_{ij}&=h^{(0)}_{ij}+\mathcal{O}(r)\,,\qquad
\gamma^{(B)}_{ir}=\mathcal{O}(r)\,,\notag\\
\beta^{(B)}_{r}&=\beta^{(0)}_r+\mathcal{O}(r)\,,\qquad
\beta^{(B)}_{i}=\beta^{(0)}_i+\mathcal{O}(r)\,.
\end{align}
It is worth noting that this is a small simplification\footnote{\label{test}If we consider the expansion
(2.4), (2.5) in \cite{Donos:2015bxe}, we first make the change
$r\to r/G^{(0)}+\dots$, which effectively sets $G^{(0)}=1$.
Next we can shift $x^i\to x^i+rf^i_{(1)}+\dots$ and 
choose the $f^i$ to eliminate the leading $r$-independent terms in $g_{ri}$.} 
of the coordinates used in 
\cite{Donos:2015bxe}. We emphasise that when we later consider theories of
gravity with higher derivatives, we will also need to keep sub-leading terms in these expansions.
Observe that we have
\begin{align}\label{textup}
\gamma^{(B)rr}&={4\pi Tr}+\mathcal{O}(r^2),\quad
\gamma^{(B)ri}=\mathcal{O}(r^2),\quad
\gamma^{(B)ij}=h^{(0)ij}+\mathcal{O}(r),\quad\nn
(\gamma^{(B)}_{D-1})^{1/2}&=\frac{\sqrt{h^{(0)}}}{(4\pi T r)^{1/2}}+\mathcal{O}(r^{1/2})\,.
\end{align}

To proceed, we now assume that the background two-forms $V$ and $W$ are well defined in the limit as $r\to 0$. 
Since the vector $\partial_r$ becomes singular as $r\to 0$ this implies that
\begin{align}\label{veedubcons}
V^{(B)ri}|_H=W^{(B)ri}|_H=0\,.
\end{align}
We will explicitly check that this condition is satisfied for the examples that we consider later.

We now define the bulk current densities via
\begin{align}\label{defjayque}
J^{i}\equiv 2(\gamma_{D-1})^{1/2}V^{ri}\,,\qquad
Q^{i}\equiv-2(\gamma_{D-1})^{1/2}W^{ri}\,,
\end{align}
By evaluating these quantities at the AdS boundary, located at $r\to\infty$, we obtain $J^{i}_\infty$ and $Q^{i}_\infty$,
which are the bulk contributions to the electric and heat currents of the dual boundary CFT. Indeed, as we explain in more detail
in appendix \ref{currcomment}, one can show that $J^{i}_\infty$ and $Q^{i}_\infty$ are the on-shell variations of the dimensionally reduced
action \eqref{eq:lower_D_functional} with respect to $\delta\beta_i^\infty=\lim_{r\to\infty}\delta \beta_i$ and
$\delta\alpha_i^\infty=\lim_{r\to\infty}\delta \alpha_i$, respectively. One then just needs to recall the parametrisation of the metric and gauge field
\eqref{eq:diff_metric_gfield_ansatz} in order to make the identification.

For the unperturbed background we write the 
bulk current densities as
\begin{align}\label{magcexp}
J^{(B)i}\equiv 2(\gamma^{(B)}_{D-1})^{1/2}V^{(B)ri}\,,\qquad
Q^{(B)i}\equiv-2(\gamma^{(B)}_{D-1})^{1/2}W^{(B)ri}\,,
\end{align}
which we notice, from \eqref{veedubcons}, vanish at the horizon.
By integrating the equations of motion \eqref{formeqmotion} in the radial direction, it is straightforward to show that we can write the background current densities at the
$r=\infty$ boundary as:
\begin{align}\label{twee}
J^{(B)i}_\infty=\partial_j M^{(B)ij}\,,\qquad
Q^{(B)i}_\infty=\partial_j M^{(B)ij}_T\,,
\end{align}
where we have defined the magnetisation densities $M^{(B)ji}$, $M^{(B)ji}_T$ as the background values
of
\begin{align}
M^{ij}\equiv\int_0^\infty dr (\gamma^{}_{D-1})^{1/2}2V^{ij}\,,\qquad
M^{ij}_T\equiv -\int_0^\infty dr (\gamma^{}_{D-1})^{1/2}2W^{ij}\,.
\end{align}
Notice that both $J^{(B)i}_\infty$ and $Q^{(B)i}_\infty$ in \eqref{twee} take the form of magnetisation
current densities
and hence are trivially spatially conserved: 
 \begin{align}
\partial_iJ^{(B)i}_\infty=0\,,\qquad \partial_iQ^{(B)i}_\infty=0\,.
\end{align}

For the DC perturbation we will also now assume that the sources only depend on the spatial directions of
the dual field theory. We therefore now take $\phi_E,\phi_T$ to be independent of the radial
coordinate and we have
\begin{align}
E=d\phi_E(x)=E_i(x) dx^i\,,\qquad \zeta =d\phi_T(x)=\zeta_i(x) dx^i\,.
\end{align}
It is illuminating to highlight some additional points concerning these sources and to do this it is helpful to focus on the particular class of
planar periodic holographic lattices. As mentioned above, by focussing on a fundamental domain, we can view the 
CFT as living on a torus. In this case we can write, for example, $\phi_T=\bar \zeta_i x^i+z(x)$ and $\zeta =\bar \zeta_i dx^i+dz(x)$,
where $\bar\zeta_i$ are constants and $z(x)$ is a periodic function. Note that the term $\bar\zeta_i dx^i$ is associated with
a constant DC thermal gradient source $\bar\zeta_i$ in the $x^i$ direction. This can be more invariantly characterised by
integrating the closed one-form $\zeta$ over a basis of one-cycles on the torus: for example, one can integrate over a period in the $x^i$ direction and then average over the length of the period.
It is worth noting that on the plane $\bar \zeta_i x^i$ is a well-defined but unbounded function, while on the torus it is a bounded
but not a well-defined function since it is not periodic. Deformations with $z(x)\ne 0$ correspond to deforming the background
but keeping the system in thermal equilibrium.
The physics of most interest is the response of the system, within a fundamental domain,
to the application of a DC source with non-zero $\bar\zeta_i$. 

For the perturbed current densities, in the presence of the sources $(\phi_T,\phi_E)$,
from \eqref{delvee}, \eqref{deldub} we have
\begin{align}\label{emcasepone}
\delta J^i&=\bar \delta J^i - \phi_T J^{(B)i}\,,\nn
\delta Q^i&= \bar \delta Q^i - 2\phi_T Q^{(B)i} - \phi_E J^{(B)i}\,,
\end{align}
where $\bar \delta J^i$, $\bar \delta Q^i$ are given by
\begin{align}\label{emcasep}
\bar \delta J^i&\equiv (\gamma_{D-1}^{(B)})^{1/2}\, 2\delta L^{ri}+\delta[(\gamma_{D-1})^{1/2}]\, 2V^{(B)ri}\,,\nn
\bar \delta Q^i&\equiv -\left((\gamma_{D-1}^{(B)})^{1/2}\, 2\delta K^{ri}+\delta[(\gamma_{D-1})^{1/2}]\, 2W^{(B)ri}\right)\,.
\end{align}
Notice 
from \eqref{emcase},\eqref{eq:GR_div_eqn}
that $\delta J^i$, $\delta Q^i$ are spatially conserved 
\begin{align}\label{conlasttime}
\partial_i\delta J^i=\partial_i\delta Q^i=0\,.
\end{align}

An important feature of \eqref{emcasepone} is the explicit appearance of $\phi_T$ and $\phi_E$ when the background
has non-vanishing magnetisation currents, $J^{(B)i},Q^{(B)i}\ne 0$. These terms are certainly physical. For example, for
periodic holographic lattices, building on the discussion above, if $\phi_T$ or $\phi_E$ are periodic functions, then 
\eqref{emcasepone} gives the perturbative change in the currents as we apply a periodic deformation to the CFT while
keeping it in thermal equilibrium. However, we are most interested in the case in which $\phi_T$ or $\phi_E$ are not periodic functions
(i.e. are not globally defined functions on the torus) and then extracting the transport currents discussed in
\cite{PhysRevB.55.2344,Blake:2015ina,Hartnoll:2007ih,Donos:2015bxe}. The transport currents of the dual field theory 
are conserved and are well-defined
in a fundamental domain, so they should not explicitly depend on $\phi_T, \phi_E$. They can be identified as follows.

We first notice that at $r\to \infty$, the contributions to the currents
$\bar \delta J^i_\infty$, $\bar \delta Q^i_\infty$ are globally defined in a fundamental domain but are not spatially conserved.
Using \eqref{emcasepone} in \eqref{conlasttime} we calculate, for example,
\begin{align}
0&=\partial_i\bar \delta J^i_\infty-J^{(B)i}_\infty\zeta_i\,,\nn
&=\partial_i\bar \delta J^i_\infty-\partial_j M^{(B)ij}\zeta_i\,,\nn
&=\partial_i\left(\bar \delta J^i_\infty +M^{(B)ij}\zeta_j\right)\,,
\end{align}
where in the last step we used $\partial_{[i}\zeta_{j]}=0$. There is 
a similar calculation for the thermal currents. We thus identify the transport currents of the dual field theory
as
\begin{align}\label{deftcs1}
\delta{\mathcal J}^i_\infty&\equiv \bar\delta{J}^i_\infty+M^{(B)ij}\zeta_j\,,\nn
\delta{\mathcal Q}^i_\infty&\equiv\bar \delta{Q}^i_\infty+2M^{(B)ij}_T\zeta_j+M^{(B)ij}E_j \,.
\end{align}
These differ from the one-point functions $\delta J^i_\infty$, $\delta Q^i_\infty$ only when there is non-vanishing magnetisation in
the background. Furthermore, $\delta{\mathcal J}^i_\infty$ and $\delta{\mathcal Q}^i_\infty$ are both globally defined on a fundamental domain and conserved:
\begin{align}
\partial_i\delta {\mathcal J}_\infty^ i=0\,,\qquad \partial_i\delta {\mathcal Q}^ i_\infty=0\,.
\end{align}

An additional perspective on these definitions is obtained as follows.
We first note that analogous to \eqref{emcasepone}, in the presence of the sources $(\phi_T,\phi_E)$,
 the perturbed magnetisations can be written
\begin{align}\label{magdefs}
\delta M^{ij}&=   \bar \delta M^{ij} - \phi_T M^{(B)ij}\,,\nn
\delta M^{ij}_T&=    \bar \delta M^{ij}_T- 2\phi_T M^{(B)ij}_T - \phi_E  M^{(B)ij}\,,
\end{align}
where $ \bar \delta M^{ij}$ and $ \bar \delta M^{ij}_T$ are both globally defined on a fundamental domain and given by
\begin{align}
\bar \delta M^{ij}&\equiv \int_0^\infty dr (\gamma_{D-1}^{(B)})^{1/2}\, 2\delta L^{ij}+\delta[(\gamma_{D-1})^{1/2}]\, 2V^{(B)ij}\,,\nn
 \bar \delta M^{ij}_T&\equiv -\int_0^\infty dr\left((\gamma_{D-1}^{(B)})^{1/2}\, 2\delta K^{ij}+\delta[(\gamma_{D-1})^{1/2}]\, 2W^{(B)ij}\right)\,.
\end{align}
We thus see that these give rise to magnetisation currents in the presence of the perturbation of the form
\begin{align}
\partial_j\delta M^{ij}&=   \partial_j\bar \delta M^{ij}- \phi_T J^{(B)i}- \zeta_jM^{(B)ij}\,,\nn
\partial_j\delta M^{ij}_T&=    \partial_j\bar \delta M^{ij}_T- 2\phi_T Q^{(B)i} - \phi_E  J^{(B)i}-2M^{(B)ij}_T\zeta_j-M^{(B)ij}E_j\,,
\end{align}
The transport current densities at $r=\infty$ can thus also be written\footnote{One might consider alternative transport currents by instead subtracting $\partial_j \delta M^{ij}$ and
$\partial_j \delta M^{ij}_T$ from $ \delta J^i_\infty$ and $\delta Q^i_\infty$, respectively (and in fact this was done in \cite{Donos:2015bxe}).
Such transport currents differ from the above definitions by the trivially conserved and globally defined magnetisation currents 
$\partial_j \bar\delta M^{ij}$, $\partial_j \bar\delta M^{ij}_T$. The definitions we use here have the property that they agree with
$\delta J^i_\infty$ and $\delta Q^i_\infty$ in the case that there are no {\it background} magnetisation currents.
It is worth emphasising that when $M^{(B)ij}=M^{(B)ij}_T=0$, in general we still have 
$\partial_j \bar\delta M^{ij}\ne 0$ and $\partial_j \bar\delta M^{ij}_T\ne 0$.}
\begin{align}\label{deftcs}
\delta{\mathcal J}^i_\infty&=\delta J^i_\infty -(\partial_j \delta M^{ij} -\partial_j \bar\delta M^{ij} )\,, \nn
\delta{\mathcal Q}^i_\infty&=\delta Q^i_\infty -(\partial_j \delta M^{ij}_T-\partial_j \bar\delta M^{ij}_T)\,,
\end{align}

\newcommand{\overbar}[1]{\mkern 2.0mu\overline{\mkern-2.0mu#1\mkern-2.0mu}\mkern 2.0mu}

With these definitions for the transport currents, 
after integrating \eqref{emcase},\eqref{eq:GR_div_eqn} in the radial direction, we deduce that the local
transport current densities at $r=\infty$ are related to the local current densities at
the horizon via
\begin{align}\label{lastcexp}
\delta{\mathcal J}^i_\infty -\partial_j \bar\delta M^{ij} =\delta J^i_H\,,\qquad
\delta{\mathcal Q}^i_\infty-\partial_j \bar\delta M^{ij}_T =\delta Q^i_H
\,,
\end{align}
where we used $\delta J^i_H=\bar\delta J^i_H$, $\delta Q^i_H=\bar \delta Q^i_H$.
We next define current flux densities $\overbar{\delta{\mathcal J}^i_\infty}, \overbar{\delta{\mathcal Q}^i_\infty}$ through a basis of $(D-3)$-dimensional cycles
on the spatial manifold on which the dual CFT lives, exactly as in \cite{Donos:2015gia,Banks:2015wha,Donos:2015bxe}. Rather than repeat the details of the definitions 
here in general, let us just note that for the special case of the periodic holographic lattices in which the spatial manifold has planar topology,
these current flux densities can be equivalently and simply defined as the zero modes of
$\delta{\mathcal J}^i_\infty$,
$\delta{\mathcal Q}^i_\infty$, which can be obtained by taking the average integral over a fundamental domain.
From \eqref{lastcexp} we then have the key result:
\begin{align}\label{result}
\overbar{\delta{\mathcal J}^i_\infty}=\overbar{\delta J^i_H}\,,\qquad
\overbar{\delta{\mathcal Q}^i_\infty}=\overbar{\delta Q^i_H}\,.
\end{align}

Having established \eqref{result}, we would like to determine how the local current densities at the horizon ${\delta J^i_H}$, ${\delta Q^i_H}$ depend on the perturbation. We make some general comments in the next subsection, based on the conditions imposed by demanding that
the perturbation is regular at the horizon. In subsequent sections, using specific theories of gravity
we show how the currents at the horizon take the form of constitutive relations for an auxiliary fluid. For the special case of
Gauss-Bonnet gravity we will show the horizon currents can be obtained by solving a generalised set of Stokes equations on the
horizon.

\subsection{Regularity of the perturbation at the horizon}
 We now examine the conditions that we need to impose in order that the perturbation \eqref{eq:forms_pert} is regular at
 the black hole horizon. To do this, near the horizon we first perform the combined coordinate and gauge transformation given by
\begin{align}\label{coordandgauge}
t&\rightarrow t\,(1+\phi_{T})\,,\notag\\
A&\rightarrow A+d\Lambda,\quad
\Lambda=-\phi_{E}\,t\,.
\end{align}
After these transformations the metric and gauge field perturbations in $D$ dimensions takes the form
\begin{align}
\delta ds^{2}&=-2H^{(B)}\,\delta H\,(dt+\alpha^{(B)})^{2}-2\,H^{(B)}{}^{2}\,(dt+\alpha^{(B)})(\delta\alpha+t\,\zeta)+\delta \gamma_{mn}\,dx^{m} dx^{n}\,,\notag\\
\delta A&=\delta A_{t}\,(dt+\alpha^{(B)})+A_{t}^{(B)}\,(\delta\alpha+t\,\zeta)+\delta\beta-t\,E\,.
\end{align}
Notice that the transformed perturbations still satisfy the covariant equations \eqref{emcase} and \eqref{eq:GR_div_eqn}.

In order to impose regular, infalling boundary conditions we define the Eddington-Finkelstein coordinate $v=t+\tfrac{\ln r}{4\pi T}$ close to the horizon. To ensure that all our fields are regular functions of $v$ and $r$ in the $r\to 0$ limit we demand that the perturbation has the following expansion
\begin{align}\label{eq:nh_pert}
\delta\alpha_{i}&=\frac{1}{4\pi T\,r }\,v_{i}+\frac{\ln r}{4\pi T}\,\zeta_{i}+\mathcal{O}(1),\qquad
\delta\alpha_{r}=-\frac{1}{4\pi T\,r }\,\delta g_{tr}^{(0)}+\mathcal{O}(1)\,,\notag\\
\delta H&=-\frac{(4\pi T  r)^{1/2}}{2} \delta g_{tt}^{(0)} +\mathcal{O}(r^{3/2}),\qquad
\delta \gamma_{rr}=\frac{1}{4\pi T r} \delta g_{rr}^{(0)}+\mathcal{O}(1)\,, \notag\\
\delta \gamma_{ri}&=- \frac{1}{4\pi T r}v_{i}+\mathcal{O}(1) ,\qquad\qquad\qquad
\delta \gamma_{ij}=\delta g_{ij}^{(0)}+2\alpha^{(0)}_{(i}v_{j)}+\mathcal{O}(r)\,,\notag\\
\delta A_{t}&=w+\mathcal{O}(r),\qquad\qquad\qquad\qquad\quad
\delta \beta_{r}=\frac{1}{4\pi T r}w+\mathcal{O}(1)\,,\notag\\
\delta \beta_{i}&=\frac{\ln r}{4\pi T}(-E_{i}+A_{t}^{(B)}\,\zeta_{i})+\delta\beta_{i}^{(0)}+\mathcal{O}(r)\,,
\end{align}
along with $\delta g_{tt}^{(0)}+\delta g_{rr}^{(0)}-2 \delta g_{tr}^{(0)}=0$.
This is the same\footnote{Note that the sign of the $tg_{tt}$ term in (4.1) of \cite{Donos:2015bxe}, should actually be a plus rather than minus.
We should also make the identification $\delta g_{ij}^{(0)}$ of \cite{Donos:2015bxe} via
 $\delta g_{ij}^{(0)}=\delta \gamma_{ij}^{(0)}-2\alpha^{(0)}_{(i}v_{j)}$.} expansion given in (4.1)-(4.3) of \cite{Donos:2015bxe}. In particular, $v_i$, $w$ and $\delta g_{tr}^{(0)}$ are 
the same quantities that entered in the Stokes equations in \cite{Donos:2015bxe}. Recalling that we are using a slightly 
different radial coordinate for the background black holes near the horizon (see footnote \ref{test}) we can identify the pressure, $p$, of \cite{Donos:2015bxe}
via $p=-(4\pi T)\delta g_{tr}^{(0)}$.
For further calculations, it is helpful to note that we have
\begin{align}\label{dgamup}
\delta \gamma^{rr}&=\mathcal{O}(r),\qquad
\delta \gamma^{ir}={v^i}+\mathcal{O}(r),\qquad
\delta \gamma^{ij}=\mathcal{O}(1),\nn
\delta[(\gamma_{D-1})^{1/2}]&= \mathcal{O}(r^{-1/2})\,.
\end{align}
with $v^i\equiv h^{(0)ij}v_j$. 

It is now straightforward to establish 
\begin{align}
\delta v_{ri}=-\frac{1}{4\pi T r}\left[E_i+\nabla_i w+a^{(0)}_t v_i\right]+a^{(0)}_t\frac{\ln r}{4\pi T}\zeta_i+\mathcal{O}(1)\,,\qquad
\delta v_{ij}=\mathcal{O}(1)\,,
\end{align}
and hence 
\begin{align}\label{delvriup}
\delta v^{ri}=-\left[E_i+\nabla_i w+a^{(0)}_t v_i+(d\beta^{(0)})^i{}_{j} v^j\right]+\mathcal{O}(r\ln r)\,.
\end{align}
Similarly,
\begin{align}
\delta w^{ri}&=-(4\pi T)^{1/2}\frac{1}{r^{1/2}}v^i+\mathcal{O}(r^{1/2})\,,\qquad
\delta w^{ij}=\mathcal{O}(r^{1/2})\,.
\end{align}
Interestingly, the sub-leading terms in these expansions are important for higher derivative theories, as we will see
later.

To obtain explicit expressions for the currents $\delta J^i_H$, $\delta Q^i_H$, we need to evaluate 
the expressions \eqref{emcasep} at the horizon. To do this we need more information on the theory of gravity that we are considering.
We will illustrate with two examples, first recovering the
Einstein-Maxwell results of \cite{Donos:2015gia,Banks:2015wha,Donos:2015bxe} in section \ref{emcasesec}, 
before moving on to higher derivative pure gravity in section \ref{gbonsec}.

\section{DC currents in Einstein-Maxwell theory}\label{emcasesec}
For the case of Einstein-Maxwell theory we can now easily recover the results
for the currents at the horizon that were obtained in \cite{Donos:2015gia,Banks:2015wha,Donos:2015bxe}.
The $D$ dimensional bulk action is given by
\begin{align}
S=\int\,d^{D}x\,\sqrt{-g}\,\left( R-V_0-\frac{1}{4}\,F^{2} \right)\,.
\end{align}
with $V_0$ constant. If we choose $V_0=-(D-1)\,(D-2)$ then 
a unit radius $AdS_D$ solves the equation of motion. The equations of motion for  
the general ansatz \eqref{eq:diff_metric_gfield_ansatz} can be obtained from
the $D-1$ dimensional action given by \eqref{eq:lower_D_functional}, \eqref{ellform} with
\begin{align}\label{eq:EM_reduced_action}
&\qquad \mathcal{L}\left(h,u,v,w,\gamma_{mn}\right)=R_{D-1}+\frac{1}{4}\,w^{2}-\frac{1}{4}\,v^{2}+\frac{1}{2}\,u^{2}-V_0\,,
\end{align}
where $R_{D-1}$ is the Ricci scalar for the $D-1$ dimensional metric on $M_{D-1}$. Notice that for this theory
$\mathcal{L}$ happens to be independent of $h$. 

We now immediately obtain
\begin{align}
\frac{\delta \mathcal{L}}{\delta v_{mn}}=-\frac{1}{2}\,v^{mn},\qquad 
\frac{\delta \mathcal{L}}{\delta w_{mn}}=\frac{1}{2}\,w^{mn}\,.
\end{align}
We thus see that the two-forms $V$, $W$ defined in \eqref{eq:form_eoms} are given by
\begin{align}
2V_{mn}&=-H v_{mn}\,,\qquad 2W_{mn}=H^2w_{mn}-HA_t v_{mn}\,,
\end{align}
where $v,w$ are defined in \eqref{eq:forms_def}. Recalling the near horizon expansions \eqref{nhorexp},
\eqref{nhorexp2} we immediately deduce that for the background black holes the two-form $V,W$ are both
well defined at the black hole horizon. In particular we find 
\begin{align}\label{veedubcons2}
V^{(B)ri}|_H=\mathcal{O}(r^{3/2}),\qquad W^{(B)ri}|_H=\mathcal{O}(r^{5/2})\,.
\end{align}
with both vanishing at the horizon, as assumed in \eqref{veedubcons}.

Turning to the perturbation, the quantities defined in \eqref{eq:L_def} and \eqref{eq:K_def2} are given by
\begin{align}\label{delLandK}
2\delta L^{mn}&=-H^{(B)}\,\delta v^{mn}-v^{(B)mn}\delta H\,,\notag\\
2\delta K^{mn}&=2H^{(B)}w^{(B)mn}\delta H+H^{(B)}{}^{2}\,\delta w^{mn}
- H^{(B)} v^{(B)mn} \delta A_t
+A^{(B)}_{t}\delta L^{mn}\,.
\end{align}
We can now evaluate the currents \eqref{emcasep} at the horizon.
For $\delta J^i_H$ we calculate as follows
\begin{align}\label{emcaseptwo}
\delta J^i_H&=\left[(\gamma_{D-1}^{(B)})^{1/2}\, 2\delta L^{ri}+\delta[(\gamma_{D-1})^{1/2}]\, 2 V^{(B)ri}\right]_H\,,\nn
&=\frac{ \sqrt{h^{(0)}}}{(4\pi T r)^{1/2}} \left[2\delta L^{ri}\right]_H\,,\nn
&=-\sqrt{h^{(0)}}[\delta v^{ri}]_H\,.
\end{align}
To get to the second line we use \eqref{veedubcons2}, \eqref{dgamup} and \eqref{textup}.
To get to the third line we use \eqref{nhorexp2}, \eqref{textup} to show that $v^{(B)ri}$ is of order $\mathcal{O}(r)$
and then use \eqref{eq:nh_pert} to show that the second term in \eqref{delLandK} does not contribute.
Finally, using \eqref{delvriup} we obtain
\begin{align}\label{emcasepfinalj}
\delta J^i_H&=\sqrt{h^{(0)}}\left[E^i+\nabla^i w+a^{(0)}_t v^i+(d\beta^{(0)})^i{}_{k} v^k\right]\,,
\end{align}
where all indices are now raised with the horizon metric $h^{(0)ij}$.

We next consider the perturbed heat current at the horizon $\delta Q^i_H$ to similarly find
\begin{align}
\delta Q^i_H&= -\left[(\gamma_{D-1}^{(B)})^{1/2}\, 2\delta K^{ri}+\delta[(\gamma_{D-1})^{1/2}]\, 2W^{(B)ri}\right]_H\,,\nn
&= - \frac{\sqrt{h^{(0)}}}{(4\pi T r)^{1/2}}\, \left[2\delta K^{ri}\right]_H\,,\nn
&= - \sqrt{h^{(0)}}(4\pi T r)^{1/2}\, \left[\delta w^{ri}\right]_H\,,
\end{align}
leading to
\begin{align}\label{emcasepfinalq}
\delta Q^i_H&=(4\pi T)\sqrt{h^{(0)}} v^i\,.
\end{align}

We have now obtained expressions for the currents at the horizon that take the form of constitutive
relations for an auxiliary fluid at the black hole. The expressions
given in \eqref{emcasepfinalj},\eqref{emcasepfinalq} are precisely the same as those that were
derived in \cite{Donos:2015gia,Banks:2015wha,Donos:2015bxe}, taking into account that we are using a slightly different radial coordinate
and hence, as noted in footnote \ref{test} above, the function $G^{(0)}$ in \cite{Donos:2015gia,Banks:2015wha,Donos:2015bxe} is set to unity.

\section{DC currents in higher derivative theories}\label{gbonsec}

For theories of gravity that involve higher derivatives
we obtain a simple modification of the two-forms $V_{mn}$ and $W_{mn}$ that were given in \eqref{eq:form_eoms}.
This is due to the fact that after dimensionally reducing on the time direction the
$(D-1)$-dimensional action \eqref{eq:lower_D_functional} will also depend on derivatives of $v$ and $w$ defined in \eqref{eq:forms_def} via \eqref{ellform}. 

For notational reasons it is convenient to introduce the operator $\mathcal{D}^{(n)}$ acting on a $(p,q)$ tensor $\Phi$ according to
\begin{align}
\mathcal{D}^{(n)}_{m_{1}\ldots m_{n}}\Phi^{\alpha_{1}\ldots \alpha_{p}}_{\beta_{1}\ldots \beta_{q}}\equiv \nabla_{m_{1}}\cdots\nabla_{m_{n}}\Phi^{\alpha_{1}\ldots \alpha_{p}}_{\beta_{1}\ldots \beta_{q}}\,.
\end{align}
Since the dimensionally reduced Lagrangian $\cal L$ given in \eqref{eq:lower_D_functional},  \eqref{ellform}
will be a function not only of $v$ and $w$ but also of $\mathcal{D}^{(s)}v$ and $\mathcal{D}^{(s)}w$, this will slightly modify
the equations of motion for $\alpha$ and $\beta$. The important point, though, is that they will have the same form given
in \eqref{formeqmotion} but now with
\begin{align}\label{genVW}
V^{mn}&=\sum_{s}(-1)^{s}\,\mathcal{D}^{(s)}_{m_{1}\ldots m_{s}}\left(H\,\frac{\delta\mathcal{L}}{\delta D^{(s)}_{m_{1}\ldots m_{s}}v_{mn}}\right)\,,\nn
W^{mn}&=H\,\sum_{s}(-1)^{s}\,\mathcal{D}^{(s)}_{m_{1}\ldots m_{s}}\,\left(H\,\frac{\delta\mathcal{L}}{\delta D^{(s)}_{m_{1}\ldots m_{s}}w_{mn}}\right)+
A_{t}\,\sum_{s}(-1)^{s}\,\mathcal{D}^{(s)}_{m_{1}\ldots m_{s}}\,\left(H\,\frac{\delta\mathcal{L}}{\delta D^{(s)}_{m_{1}\ldots m_{s}}v_{mn}}\right)\,.
\end{align}

As a consequence the results in section \ref{gendisc} that we derived for theories with two derivatives 
all generalise to theories with higher derivative after performing the following replacements:
\begin{align}\label{repls}
\frac{\delta\mathcal{L}}{\delta v_{mn}}&\rightarrow  (\delta_{v}\mathcal{L})^{mn}\equiv H^{-1}  \sum_{s}(-1)^{s}\,\mathcal{D}^{(s)}_{m_{1}\ldots m_{s}}\left(H\,\frac{\delta\mathcal{L}}{\delta D^{(s)}_{m_{1}\ldots m_{s}}v_{mn}}\right)\,,\nn
\frac{\delta\mathcal{L}}{\delta w_{mn}}&\rightarrow (\delta_{w}\mathcal{L})^{mn}\equiv H^{-1} \sum_{s}(-1)^{s}\,\mathcal{D}^{(s)}_{m_{1}\ldots m_{s}}\,\left(H\,\frac{\delta\mathcal{L}}{\delta D^{(s)}_{m_{1}\ldots m_{s}}w_{mn}}\right)\,.
\end{align}

\subsection{Higher derivative gravity}

For the remainder of the paper we are going to focus on the general class of theories of pure gravity in $D$ spacetime dimensions
with Lagrangian of the form
\begin{align}\label{hdact}
\mathcal{L}&=R-V_0+c_{1}\mathcal{L}_1+c_{2}\,\mathcal{L}_2+c_{3}\mathcal{L}_3\,,
\end{align}
where $V_0$ and $c_i$ are constants with
\begin{align}
\mathcal{L}_1=R_{\mu_{1}\mu_{2}\mu_{3}\mu_{4}}R^{\mu_{1}\mu_{2}\mu_{3}\mu_{4}},\qquad
\mathcal{L}_2=R_{\mu_{1}\mu_{2}}R^{\mu_{1}\mu_{2}}\,,\qquad
\mathcal{L}_3=R^2\,.
\end{align}
Within the context of string theory, one is interested in solutions to the leading order equations with 
corrections that are perturbative in the $c_a$ which will be of order $\alpha^{\prime}$. The case of Gauss-Bonnet gravity in $D\ge 5$ spacetime dimensions corresponds to $c_{1}=c_{3}=-\frac{1}{4}\,c_{2}\equiv\tilde \alpha$. We now aim to calculate the 
perturbed heat current at the horizon after switching on a DC source, parametrised by a closed one-form
$\zeta$.

We first carry out the dimensional reduction on the time coordinate\footnote{A Kaluza-Klein reduction on a spatial coordinate is analysed in \cite{Baskal:2010sv}.}
to obtain the $(D-1)$-dimensional theory. We introduce the obvious orthonormal frame
$(e^0,e^{\hat m})$
associated with the $D$-dimensional spacetime metric given in
\eqref{eq:diff_metric_gfield_ansatz}, with
$e^0=H( dt+\alpha)$ and $e^{\hat m}e^{\hat m}=ds^{2}\left(M_{D-1} \right)$.
We calculate the
various components of the Riemann tensor of the $D$-dimensional metric to get
\begin{align}\label{riemkk}
R_{\hat m\hat n\hat p\hat q}&=\bar R_{\hat m\hat n\hat p\hat q}+\frac{1}{2}w_{\hat m\hat n}w_{\hat p\hat q}-\frac{1}{2}w_{\hat m[\hat p}w_{\hat q]\hat n}\equiv\Sigma^{(4)}_{\hat m\hat n\hat p\hat q}\,,\nn
R_{\hat m0\hat n\hat p}&=\frac{1}{2}\nabla_{\hat m} w_{\hat n\hat p}+\frac{1}{2}h_{\hat m}w_{\hat n\hat p}-h_{[\hat n}w_{\hat p]\hat m}\equiv\Sigma^{(3)}_{\hat m\hat n\hat p}\,,\nn
R_{\hat m0\hat n0}&=\nabla_{\hat m}h_{\hat n}+h_{\hat m}h_{\hat n}-\frac{1}{4}w_{\hat m\hat p} w^{\hat p}{}_{\hat n}\equiv\Sigma^{(2)}_{\hat m\hat n}\,,
\end{align}
where $\bar R_{\hat m\hat n\hat p\hat q}$ are the components of the Riemann tensor for the $(D-1)$-dimensional metric $\gamma_{D-1}$ in the orthonormal frame.
It is worth noting the Bianchi identity that arises from the definition of $w$ takes the form $\nabla_{[\hat m} w_{\hat n\hat p]}=h_{[\hat m}w_{\hat n\hat p]}$.
For the components of the Ricci tensor in the orthonormal frame we have
\begin{align}\label{ricexps}
R_{\hat m \hat n}&=\bar R_{\hat m \hat n}+\frac{1}{2}w^2_{\hat m\hat n}-\nabla_{\hat m} h_{\hat n}- h_{\hat m}h_{\hat n}\equiv\Lambda^{(2)}_{\hat m\hat n}\,,\nn
R_{\hat m0}&=h^{\hat n}w_{\hat n\hat m}+\frac{1}{2}\nabla^{\hat n} w_{\hat n\hat m}\equiv\Lambda^{(1)}_{\hat m}\,,\nn
R_{00}&=\nabla_{\hat m} h^{\hat m}+h^2+\frac{1}{4}w^2\equiv\Lambda^{(0)}\,,
\end{align}
where $w^2=w_{m n}w^{mn}$ and $h^2=h_{m}h^{m}$. Finally, the Ricci scalar is given by
\begin{align}\label{rsexps}
R=\bar R+\frac{1}{4}w^2-2\nabla_m h^m-2h^2\,.
\end{align}

We thus have
\begin{align}
\mathcal{L}^{(1)}&=\Sigma^{(4)}_{mnpq}\Sigma^{(4)}{}^{mnpq}-4\,\Sigma^{(3)}_{mnp}\Sigma^{(3)}{}^{mnp}+4\,\Sigma^{(2)}_{mn}\Sigma^{(2)}{}^{mn}\,,\notag\\
\mathcal{L}^{(2)}&=\Lambda^{(2)}_{mn}\Lambda^{(2)}{}^{mn}-2\,\Lambda^{(1)}_{m}\Lambda^{(1)}{}^{m}+\Lambda^{(0)}{}^{2}\,,\nn
\mathcal{L}^{(3)}&=R^2\,.
\end{align}
Using the definitions \eqref{repls} we compute
\begin{align}\label{defgenW}
(\delta_{w}\mathcal{L}^{(1)})^{mn}&=2\left(\Sigma^{(4)}{}^{mn}{}_{pq}+\,\Sigma^{(4)}{}^{\left[m\right.}{}_{p}{}^{\left. n\right]}{}_{q} \right)\,w^{pq}\notag\\
&\quad -4\,h_{p}\,\left(\Sigma^{(3)}{}^{pmn}-2\,\Sigma^{(3)}{}^{\left[ mn \right]\,p} \right)+4H^{-1}\,\nabla_{p}\left(H\,\Sigma^{(3)}{}^{pmn}\right) -4\,\Sigma^{(2)}{}^{p\left[ m\right.}w^{\left. n\right]}{}_{p}\,,\notag\\
(\delta_{w}\mathcal{L}^{(2)})^{mn}&=-2\,\Lambda^{(2)}{}^{p\,\left[m\right.}w^{\left. n\right]}{}_{p}+4\,\Lambda^{(1)}{}^{\left[m\right.}h^{\left. n\right]}+2H^{-1}\,\nabla^{\left[m \right.}H\,\Lambda^{(1)}{}^{\left. n\right]}+\Lambda^{(0)}\,w^{mn}\,,\notag\\
(\delta_{w}\mathcal{L}^{(3)})^{mn}&=R\,w^{mn}\,.
\end{align}

We also record here that for this theory instead of \eqref{eq:form_eoms} and \eqref{eq:K_def2} we have
\begin{align}\label{wkay}
W^{mn}&=H^2(\delta_w\mathcal{L})^{mn}\,,\nn
\delta K^{mn}&\equiv \,2W^{(B)mn}H^{(B)-1}\,\delta H+H^{(B)}{}^{2}\,\delta \left[\delta_w\mathcal{L}\right]^{mn}\,.
\end{align}

\subsection{Heat current in a static background}

For simplicity we now focus on static background solutions and set
$\alpha^{(B)}=0$. For this higher derivative theory, it turns out that we need to keep sub-leading terms
in the expansion of the background fields about the black hole horizon.
We thus take, for the background as $r\to 0$,
\begin{align}\label{nhorexpa}
&\gamma_{rr}^{(B)}=\frac{1}{4\pi T\,r}\,\left( 1+\gamma_{rr}^{(1)}\,r\right)+\mathcal{O}(r)\,,\quad
\gamma^{(B)}_{ij}=h^{(0)}_{ij}+h^{(1)}_{ij}\,r+\mathcal{O}(r^{2})\,,\quad
\gamma^{(B)}_{ir}=\mathcal{O}(r)\,,\nn
&H^{(B)}{}^{2}= 4\pi T\,r\,\left(1+2\,H^{(1)}\,r \right)+\mathcal{O}(r^{3})\,.
\end{align}
There is still some residual freedom in our choice of the radial coordinate.
If we make the shift $r\to r+r^2f(x)$ then this leads to the same fall-offs but with 
$H^{(1)}\to H^{(1)}+f/2$ and $\gamma^{(1)}_{rr}\to
\gamma^{(1)}_{rr}+3f$. As a consequence, 
$H^{(1)}$ and $\gamma^{(1)}_{rr}$ can only appear in our final expressions
in the combination
$3H^{(1)}-\gamma^{(1)}_{rr}/2$, as we shall see.
An important point is that while it is is necessary to keep sub-leading terms in the expansion of the background, we find that
it is not necessary to include the sub-leading terms in the expansion of the perturbation
\eqref{eq:nh_pert} at the horizon.

We are now ready to calculate $\delta Q^i$ at the horizon. From the definition \eqref{emcasep} we have
\begin{align}\label{defdelqueue}
\delta Q^i_H&= -\left[(\gamma_{D-1}^{(B)})^{1/2}\, 2\delta K^{ri}+\delta[(\gamma_{D-1})^{1/2}]\, 2W^{(B)ri}\right]_H\,.
\end{align}
We are considering static backgrounds with $\alpha=w=0$. From \eqref{riemkk},\eqref{ricexps}
we deduce that in the background we have $\Sigma^{(3)}=\Lambda^{(1)}=0$ and hence from
\eqref{defgenW},\eqref{wkay} we deduce that $W^{(B)mn}=0$.
We thus have
\begin{align}
\delta Q^i_H&=-\left[2\sqrt{\gamma}H^{(B)2}\delta[\delta_w\mathcal{L}]^{ri}\right]_H\,.
\end{align}
The contribution to this expression from the two-derivative part of the action in \eqref{hdact}
is the same as before and
so we have
\begin{align}\label{cqsa}
\delta Q^i_H&=(4\pi T)\sqrt{h^{(0)}} v^i
+c_1\delta Q_H^{(1)}{}^{i}+c_2\delta Q_H^{(2)}{}^{i}+c_3\delta Q_H^{(3)}{}^{i}\,,
\end{align}
where
\begin{align}
\frac{1}{4\pi T \sqrt{h_{(0)}}}\delta Q_H^{(a)}{}^{i}\equiv& 
-\frac{2r^{1/2}}{(4\pi T)^{1/2}}\,\left[\delta(\delta_{w}\mathcal{L}^{(a)})^{ri}\right]_H\,.
\end{align}

After some extensive calculations, which we describe in appendix \ref{details}, we can obtain the contributions from the
higher-derivative terms in the action. For the Riemann squared part of the action we get
\begin{align}\label{qonefe}
\frac{1}{4\pi T \sqrt{h_{(0)}}}\delta Q_H^{(1)}{}^{i}=&
4\,\nabla_{j}\nabla^{[j}v^{i]}+4R_{(0)}^{ij}v_{j}+4\nabla^i\nabla_jv^j\nn
&+2(4\pi T)\left[ 2\left(\zeta^{i}+\nabla^{i}\delta g_{tr}^{(0)}\right)-\left(3\,H^{(1)}-\frac{1}{2}\gamma_{rr}^{(0)} \right) v^{i}\right]\,.
\end{align}
For the Ricci squared part of the action we get 
\begin{align}\label{qtwofe}
&\frac{1}{4\pi T \sqrt{h_{(0)}}}\delta Q_H^{(2)}{}^{i}=
2\,\nabla_{j}\nabla^{[j}v^{i]}+2R_{(0)}^{ij}v_{j}+\nabla^i\nabla_jv^j\nn
&\qquad\qquad+(4\pi T)\left[\left( \zeta^{i}+\nabla^{i} \delta g_{tr}^{(0)}\right) 
-\left(3\,H^{(1)}-\frac{1}{2}\gamma_{rr}^{(1)} +\frac{1}{2}h^{(1)}{}^j{}_j\right)v^i           \right]    \,.
\end{align}
Finally for the Ricci scalar squared part of the action we get 
\begin{align}\label{qthreefe}
\frac{1}{4\pi T \sqrt{h_{(0)}}}\delta Q_H^{(3)}{}^{i}=
&2R^{(0)}v^i-2\,(4\pi T)
\left(3H^{(1)}-\frac{1}{2}\gamma^{(1)}_{rr}+h^{(1)}{}^j{}_j\right)v^{i}\,.
\end{align}

A number of comments are now in order.
Firstly, the expressions for the heat current above depend on the horizon metric as well as the sub-leading corrections to the background black hole solution that are parametrised by $3H^{(1)}-\frac{1}{2}\gamma^{(1)}_{rr}$ and $h^{(1)}_{ij}$. In principle these sub-leading corrections can be
expressed in terms of the geometry of the horizon after using the equations of motion. We will not carry
out this calculation here, but instead below we will analyse the results that one obtains by considering the higher-derivative corrections to the action to be perturbatively small. 

Second, from our general analysis in section \ref{gendisc} we know that we have $\partial_i \delta Q^{(a)i}=0$
when we use the equations of motion. In this regard, we note that $\nabla_i\nabla_j\nabla^{[j}v^{i]}=0$.

Third, we see that the expressions for the heat current are expressed in terms of the perturbation via $v^i$, as we saw for two-derivative theories of gravity, as well as $\zeta^{i}$ and $\nabla_i\delta g_{tr}^{(0)}$. For two-derivative theories of gravity, 
$\nabla_i\delta g_{tr}^{(0)}$ appeared\footnote{In comparing with e.g. equation (4.7) in \cite{Donos:2015bxe} on should take into account that we are using a different radial variable.} in the Stokes equations, via the gradient of a pressure term $\nabla_ip=-(4\pi T)\nabla_i\delta g_{tr}^{(0)}$. Thus we can continue to interpret the expressions for the heat current as constitutive relations
for an auxiliary fluid on the horizon.

Finally, for the special case of Gauss-Bonnet gravity in $D\ge 5$ spacetime dimensions,
notice that when we combine the currents with
$c_{1}=c_{3}=-\frac{1}{4}\,c_{2}\equiv \tilde\alpha$ there is a cancellation of many terms and 
we find the simple expression
\begin{align}\label{fchere}
\delta Q^{i}_H
={4\pi T \sqrt{h_{(0)}}}\left[v^i-4\tilde\alpha\left(\nabla_{j}\nabla^{[j}v^{i]}
+G_{(0)k}^{\,\,\,\,\,i}v^{k}\right)\right]\,,
\end{align}
where we have defined the Einstein tensor for the background $G_{(0)}^{ij}=R_{(0)}^{ij}-\frac{1}{2}R_{(0)}\,h_{(0)}^{ij}$.
The conservation of this current is equivalent to
\begin{align}\label{gbconn}
\nabla_iv^i=4\tilde\alpha G^{\,\,ij}_{(0)} \nabla_i v_j\,,
\end{align}
where in the above $\nabla$ is the covariant derivative associated with $h_{(0)}^{ij}$.
Our general analysis implies that \eqref{gbconn} must follow from the Gauss-Bonnet equations of motion and we will show that this is true in the next section.

Returning to theories of gravity with general $c_i$, we now derive expressions for the currents when the higher-derivative corrections
are perturbatively small. At zeroth order in the corrections we can use the leading Einstein equations in $D$
spacetime dimensions, $R_{\mu\nu}=DV_0/(D-2)g_{\mu\nu}$, in order to obtain expressions for the leading order
corrections to the background expansions near the horizon. Using the decompositions \eqref{ricexps},\eqref{rsexps}
combined with \eqref{tink1},\eqref{tink2} and \eqref{nhorexpa} we easily obtain
\begin{align}\label{eq:alpha_pert}
3H^{(1)}-\frac{1}{2}\gamma_{r}^{(1)}&=\frac{1}{4\pi T}\,\left(-R_{(0)}+\frac{D-4}{D-2}V_0 \right)\,,\nn
h^{(1)}_{ij}&=\frac{2}{4\pi T}\left( R_{(0)}{}_{ij}-\frac{1}{D-2} V_0h_{(0)}{}_{ij}\right)\,.
\end{align}
After substituting these leading order expressions into $\delta Q_H^{(a)}{}^{i}$ we obtain
\begin{align}
\frac{1}{4\pi T \sqrt{h_{(0)}}}\delta Q_H^{(1)}{}^{i}\approx&\,\,4\,\nabla_{j}\nabla^{(j}v^{i)}
+\left( 2 R_{(0)}-2\frac{D-4}{D-2}V_0\right)v^{i}
+4\,(4\pi T)\left(\zeta^{i}+\nabla^{i}\delta g_{tr}^{(0)}\right)\,,\nn
\frac{1}{4\pi T \sqrt{h_{(0)}}}\delta Q_H^{(2)}{}^{i}\approx&\,\,\nabla^2 v^i+R_{(0)}^{ij}v_{j}+\frac{2V_0}{D-2} v^{i}
+\,(4\pi T)\left(\zeta^{i}+\nabla^{i}\delta g_{tr}^{(0)}\right)\,,\nn
\frac{1}{4\pi T \sqrt{h_{(0)}}}\delta Q_H^{(3)}{}^{i}\approx&\,\,\frac{2D}{D-2}{V_0}v^{i}\,,
\end{align}
where we used the identity
\begin{align}\label{ricid}
\nabla_{j}\nabla^{[j}v^{i]}=\nabla_{j}\nabla^{(j}v^{i)}-\nabla^{i}\nabla_{j}v^{j}-R_{(0)}^{ij}v_{j}\,.
\end{align}
The expressions for $\delta Q_H^{(a)}{}^{i}$ now only depend on the intrinsic geometry of the black hole horizon, as well as the
perturbation.

In the context of holography, following the discussion in section \ref{currcomment},
 the current flux densities at the horizon, $\delta \bar Q^i_H$, are identified with the renormalised
 transport\footnote{Note that since we have assumed that the background geometry is static with $\alpha^{(B)}=0$, there are no magnetisation currents and the transport currents are the same as
the currents.} current flux densities of the dual field theory. Note that by treating the higher derivative terms
perturbatively allows one to consider the total on-shell action, including Gibbons-Hawking and counter terms,  to be still a functional of the boundary metric (for a related discussion see \cite{Cremonini:2009ih}).

In the next section, for the special case of Gauss-Bonnet gravity, we show how the local current
densities at the horizon, $\delta Q^i_H$, can be obtained by solving a higher derivative generalisation
of the Stokes equations.

\section{Gauss-Bonnet and generalised Stokes equations}\label{gbns}

In this section we will consider Gauss-Bonnet gravity in $D\ge 5$ spacetime dimensions. 
Once again, for simplicity, we will focus on static background black hole spacetimes.
We will  use a radial Hamiltonian formalism and evaluate the momentum and Hamiltonian constraints on the horizon. This will enable us to derive an expression for the local heat current density on the horizon. In fact we will obtain an expression that differs by a magnetisation term from that given in \eqref{fchere}, for reasons we
will explain later. In addition we will also obtain a higher derivative version of the Stokes equations for the auxiliary fluid living on the horizon, generalising the Stokes equations found in
\cite{Donos:2015gia}. In particular, we will obtain a closed set of equations for a subset of the perturbation at the horizon which can be solved to obtain the local heat current density on the horizon. In turn, via \eqref{result}, by then evaluating the current flux density on the horizon we can obtain the transport current flux density on the boundary.

\subsection{Radial Hamiltonian formulation}
To obtain the radial Hamiltonian formulation of Gauss-Bonnet gravity, we essentially follow \cite{Liu:2008zf}, 
who adapted the results of \cite{Teitelboim:1987zz}.
We start by writing the bulk action as
\begin{align}\label{hdact2}
S&=\int\! d^{D}\!x
\sqrt{-g}\left[
R-V_0+ \tilde{\alpha}\left(R_{\mu\nu\rho\sigma}
R^{\mu\nu\rho\sigma}-4R_{\mu\nu}R^{\mu\nu}+R^2\right)\right]\,,
\end{align}
where we have again set $16\pi G=1$ for convenience. 
If we set $V_0=-(D-1)\,(D-2)$ then 
a unit radius $AdS_D$ solves the equation of motion when $\tilde\alpha=0$.
We write the spacetime coordinates as $y^\mu=(r,x^a)$ where 
\begin{align}
x^a=(t,x^i)\,,
\end{align}
are the coordinates for the dual field theory.
We perform a radial decomposition of the bulk metric in a standard way, writing
\begin{align}
ds^2&=g_{\mu\nu}dy^\mu dy^\nu=N^2dr^2+\s_{ab}(dx^a+N^adr)(dx^b+N^bdr)\,.
\end{align}
The normal vector to surfaces of constant $r$ has components $n^\mu=N^{-1}(1,-N^a)$, while
$n_\mu=N(1,0)$. The induced metric on the surfaces of constant $r$ is given by $\sigma_{\mu\nu}=g_{\mu\nu}-n_\mu n_\nu$ and has non-vanishing components $\sigma_{ab}$.
The extrinsic curvature is defined as $K_{\mu\nu}=\tfrac{1}{2}{\cal L}_n \s_{\mu\nu}=\sigma_\mu^\rho\nabla_\rho n_\nu$ and
has non-vanishing components given by 
\begin{align}
K_{ab}=\frac{1}{2N}(\partial_r\s_{ab}-D_aN_b-D_bN_a)\,,
\end{align}
where $N_a=\s_{ab}N^b$.

The bulk action is supplemented by Gibbons-Hawking type terms given by \cite{Myers:1987yn,Teitelboim:1987zz}
\begin{align}
S_{GH}=-2\int _{\partial {\cal M}}\! d^{D-1}x \sqrt{-\s}\left[K -\tilde\alpha\left( 4 G_{ab}K^{ab}-\tfrac{2}{3}\left( K^3-3K K_{ab}K^{ab}+2K_a^bK_b^cK_cK^a \right)\right) \right]\,,
\end{align}
where $G_{ab}$ is the Einstein tensor for $\sigma_{ab}$, and
ensures that the on-shell action is a functional of the boundary metric. It is also supplemented
by boundary counter terms, which we shall not need here, but are discussed in
\cite{Brihaye:2008kh,Astefanesei:2008wz,Liu:2008zf,Cremonini:2009ih} (the regularised holographic stress tensor is discussed in
\cite{Astefanesei:2008wz}).

After dropping a total derivative, the bulk action can be rewritten as
\begin{align}\label{gbaction}
S_{GB} =& \int d^D x \sqrt{-\s} N \left(\mathcal{R}-V_0+ K^2- K_{ab}K^{ab} \right)\nn
+&\tilde \alpha \int d^D x \sqrt{-\s}N\Bigg(\left[\mathcal{R}+ K^2- K_{ab}K^{ab}\right]^2 - 4\left[\mathcal{R}_{ab} + KK_{ab} -K_{ac}K^c_{b} \right]^2\nn 
&\qquad\qquad
+ \left[\mathcal{R}_{abcd}+ K_{ac}K_{bd}- K_{ad}K_{bc}\right]^2 - \frac{4}{3}K^4 + 8 K^2K_{ab}K^{ab}\nn
&\qquad\qquad
 -\frac{32}{3}KK^b_a K^c_bK^a_c -4\left[K_{ab}K^{ab} \right]^2+8 K^b_a K^c_b K^d_c K^a_d\Bigg)\,,
\end{align}
where $\mathcal{R}_{abcd}$, $\mathcal{R}_{ab}$ and $\mathcal{R}$, are the Riemann tensor, Ricci tensor and Ricci
scalar associated with $\sigma_{ab}$.

The conjugate momenta, which are densities, are defined by
\begin{align}
\pi^{ab}=\frac{\delta S_{GB}}{\delta\dot \sigma_{ab}}=\frac{1}{2N}\frac{\delta S_{GB}}{\delta K_{ab}}\,.
\end{align}
Explicitly we have
\begin{align}
\frac{1}{\sqrt{-\s}}{\pi}^b_a&=K\delta^b_a-K^b_a+\tilde\alpha\frac{1}{\sqrt{-\s}}{\pi_{GB}}^b_a\,,
\end{align}
where:
\begin{align}
\frac{1}{\sqrt{-\s}}{\pi_{GB}}^b_a=&2K^b_a(K^2-K^c_dK^d_c-\mathcal{R})\cr
&-4K\mathcal{R}^b_a+4K^c_a\mathcal{R}_c^b+4K_c^b\mathcal{R}^c_a +4K^{cd}{\mathcal{R}^b}_{cad} -4KK_c^bK^c_a+4K_c^bK^c_dK^d_a\cr
&+\delta^b_a(2K\mathcal{R}-\frac{2}{3}K^3+2KK_c^dK^c_d-4K_c^d\mathcal{R}^c_d-\frac{4}{3}K_c^dK^c_eK_d^e)\,.
\end{align}

The Hamiltonian density is defined as $\mathcal{H}=\pi^{ab}\dot\sigma_{ab}-\sqrt{-g}\mathcal{L}$. 
After dropping a total derivative $2D_a(N_b \pi^{ab})$,  we find that $\mathcal{H}$
is a sum of constraints:
\begin{align}
\mathcal{H}=NH+N_aH^a\,,
\end{align}
where
\begin{align}\label{hcon}
\frac{1}{\sqrt{-\s}}H=V_0-&\mathcal{R}+K^2-K^a_bK^b_a-\tilde\alpha\Big[(\mathcal{R}-K^2+K^a_bK^b_a)^2\nn&-4(\mathcal{R}_{ab}-KK_{ab}+K^c_aK_{bc})^2+(\mathcal{R}_{abcd}-K_{ac}K_{bd}+K_{ad}K_{bc})^2\Big]\,,
\end{align}
and
\begin{align}\label{momcon}
\frac{1}{\sqrt{-\s}}H^a=-2D_b\left(\frac{1}{\sqrt{-\s}}{\pi}^{ba} \right)\,.
\end{align}
Note that we will not need to express $H$ in terms of the canonical momentum (this is done perturbatively in $\tilde\alpha$ in \cite{Liu:2008zf}).

\subsection{Evaluating the momenta and constraints at the horizon}
We want to evaluate the conjugate momenta as well as the momentum and Hamiltonian constraints on a surface of constant $r$ near
the horizon, and then take the limit $r\to 0$. Several details are presented in appendix \ref{conGB}.
We first note that for the background (i.e. unperturbed metric), at the horizon we have $\pi^{(B)}_H{}^t_t=\pi^{(B)}_H{}^i_t=\pi^{(B)}_H{}^t_i=0$
and
\begin{align}\label{backstress}
2\pi^{(B)}_H{}^i_j=(4\pi T)\sqrt {h_{(0)}}(\delta^i_j-4\tilde\alpha {G}^i_{(0)j})\,,
\end{align}
where ${G}_{(0)ij}$ is the Einstein tensor associated  with the horizon metric $h_{(0)ij}$. 

We next note that for the perturbed metric we have
\begin{align}\label{localcurq}
\delta\tilde Q_H^i\equiv -2\pi_H{}^i_t=2\pi^{(B)}_H{}^i_j v^j\,.
\end{align}
We see that $\delta\tilde Q_H^i$ differs from the expression for $\delta Q^{i}_H$ that we derived in \eqref{fchere}, 
by a magnetisation current piece, for reasons we explain later. Expressions for other background components can be found in appendix \ref{conGB}.

We next consider the momentum constraints $H^a=0$ with $H^a$ given \eqref{momcon}
Evaluating the time component $H_t=0$ on the horizon we find
the conservation condition $\partial_i \delta\tilde Q_H^i=0$ which is equivalent to
\begin{align}\label{incompresstext}
\nabla_i(\pi^{(B)}_H{}^i_j v^j)=\pi^{(B)}_H{}^i_j {\nabla}_iv^j=0\,,
\end{align}
where here $\nabla$ is the covariant derivative associated 
with the horizon metric $h_{(0)ij}$. In fact this is the same condition as $\partial_i \delta Q_H^i=0$ that we mentioned earlier in \eqref{gbconn}.
It can also be shown that imposing the Hamiltonian constraint $H=0$ at the horizon, where $H$ is given in \eqref{hcon},
gives rise to exactly the same condition as
\eqref{incompresstext}.

We next evaluate the space component of the momentum constraint $H_j=0$ on the horizon to obtain the Stokes equations
\begin{align}\label{nsgeneq}
&0=-{\nabla}^k{\nabla}_{(j}v_{k)}-(4\pi T\zeta_i-{\nabla}_ip)(\frac{1}{2}\delta_j^i-2\tilde\alpha{G}_{(0)}{}^i_j)\nn
&+\tilde\alpha{\nabla}_i\Bigg[-4{\nabla}_kv^k{R}_{(0)}{}^i_j  +4{\nabla}_{(j}v_{k)}{G}_{(0)}^{ik}+4{\nabla}^{(k}v^{i)}{R}^{(0)}_{jk}+4{\nabla}^{(k}v^{l)}{\,{R}^i}_{(0)kjl}\nn
&+4\pi T\left(h^{ik}_{(0)}{\nabla}_{(j}v_{k)}{h^l_l}^{(1)}+{h^i_j}^{(1)}{\nabla}_kv^k-{\nabla}^{(i}v^{k)}{h_{jk}}^{(1)}-{h^{ik}}^{(1)}{\nabla}_{(j}v_{k)}+\delta^i_j(-{h_k^k}^{(1)}{\nabla}_lv^l +{h^{kl}}^{(1)}{\nabla}_kv_l)\right)
\Bigg]\,.
\end{align}
Notice that these equations depend on the intrinsic geometry of the metric, associated with the metric 
$h_{(0)ij}$, as well as the sub-leading piece $h_{(1)ij}$ in the expansion of the background at the horizon given in
\eqref{nhorexpa}. In principle, the terms involving $h_{(1)ij}$ could be related to $h_{(0)ij}$ using the full equations of motion.
We will not investigate this here as we are most interested in working perturbatively in $\tilde\alpha$.
Using \eqref{eq:alpha_pert}, that \eqref{incompresstext} implies $\nabla_i v^i$ is order $\tilde \alpha$, as well as $V_0=-(D-1)(D-2)$ we find that the Stokes equations \eqref{nsgeneq} can be written
compactly as
\begin{align}\label{nsspecial}
-2\nabla^i\left(S_{ij}^{kl}\nabla_k v_l\right)=&\frac{2\pi^{(B)}_H{}^i_j}{\sqrt {h_{(0)}}}\left(\zeta_i-\frac{\nabla_i p}{4\pi T}\right)\,,
\end{align}
where $\pi^{(B)}_H{}^i_j$ is given in \eqref{backstress}
and $S^{kl}_{ij}=S^{(kl)}_{(ij)}$ is given by  
\begin{align}\label{expforess}
S^{kl}_{ij}=\left[1-\tilde\alpha 2(D-4)(D-1)\right]\delta^{(k}_i\delta^{l)}_j
-\tilde\alpha\left[2h^{(0)}_{ij}R_{(0)}^{kl}+4\delta^{(k}_{(i}R_{(0)}{}^{l)}_{j)}+4R_{(0)}{}_i{}^{(k}{}_j{}^{l)}\right]\,.
\end{align}

Equations \eqref{incompresstext} and \eqref{nsspecial} are the main results of this section.
A number of comments are in order. Firstly, \eqref{incompresstext} and \eqref{nsspecial} depend 
only upon $v_i$, $p$, $\zeta_i$ and background quantities. As such, for a fixed source $\zeta$,
they give a closed set of equations which can be solved for $v_i$, $p$. In turn this gives
the local conserved heat current density on the horizon via \eqref{localcurq}.
By then evaluating the current flux density on the horizon we can obtain the transport current flux density on the boundary, via \eqref{result},
and hence the thermal DC conductivity. We have thus successfully generalised the main results of 
\cite{Donos:2015gia,Banks:2015wha,Donos:2015bxe} to Gauss-Bonnet gravity.

Second, it is interesting to point out that for the special case of the homogeneous black brane solution of \cite{Cai:2001dz}, with flat horizon,
the shear viscosity for Gauss-Bonnet theory was calculated in \cite{Brigante:2007nu} and the result was given by
\begin{align}\label{shear}
4\pi\frac{\eta}{s}=1-\tilde \alpha 2(D-4)(D-1)\,.
\end{align}
Note that that this is precisely the same coefficient appearing in the first term in \eqref{expforess}.
This can be understood as follows, generalising the discussion of \cite{Banks:2016krz} in the context of ordinary two-derivative Einstein gravity. 
Imagine we calculate the DC conductivity in the hydrodynamic limit
with $\epsilon=k/T<<1$, where $k$ is the largest wavelength of the background holographic lattice, but we keep $\tilde\alpha/T^2$ corrections.
For simplicity, we also assume that the holographic lattice has $g_{tt}^\infty=-1$. In the limit $\epsilon<<1$
the holographic lattice black hole solution will be the Gauss-Bonnet black brane solution of \cite{Cai:2001dz}, but with spatial sections given
by $g_{ij}^\infty$. The DC conductivity is still obtained using the Stokes equations 
\eqref{nsspecial}, but now with the horizon metric proportional to $g_{ij}^\infty$. On the other hand, one should also be able to obtain this result
using a fluid gravity approach for studying CFTs on a curved manifold with metric $g_{ab}^\infty$. Although this has not been worked
out in detail as far as we know, the analysis for Gauss-Bonnet theory should closely follow the fluid-gravity formalism of two-derivative gravity \cite{Bhattacharyya:2008mz}. 
In particular, the shear viscosity
as in \eqref{shear} will appear as a first order transport coefficient in the fluid equations
(see \cite{Grozdanov:2016fkt} for a recent discussion). Finally, as in the  analysis of \cite{Banks:2016krz}, the DC perturbation can be studied 
within the fluid-gravity formalism using a suitable fluid flow and this will lead to a system of Stokes equations with shear viscosity 
\eqref{shear}.

Third, we now explain the origin of the difference between the $-2\pi_H{}^i_t$ and the expression for
$\delta Q^{i}_H$ that we derived in \eqref{fchere}. In moving from the action \eqref{hdact2} to
the action \eqref{gbaction} we dropped a total derivative. This will not modify the bulk equations of motion. However, it
can give a contribution to the current at the horizon and, since the bulk equations of motion are not modified, the extra contribution
should be a magnetisation current in order that it is trivially conserved. It is possible to explicitly check this in detail, but we shall not do so here.

Fourth, we notice that we can obtain the system of equations \eqref{incompresstext}, \eqref{nsspecial}
by varying the following Lagrangian
\begin{align}
L=\int d^{D-2}x\Big[-\sqrt{h_{(0)}}\nabla^iv^jS^{kl}_{ij}\nabla_k v_l+2\pi^{(B)}_H{}^i_j\left(
v^j \zeta_i
+\frac{p}{4\pi T}\nabla_i v^j
\right)\Big]\,,
\end{align}
with respect to $v^i$ and $p$. It is also interesting to note that
we can obtain the local heat current density at the horizon, $\pi_H{}^i_t$, given in
\eqref{localcurq}, if we vary with respect to $\zeta^i$:
\begin{align}
\delta\tilde Q_H^i=\frac{\delta L}{\delta \zeta^i}\,.
\end{align}
By taking an additional derivative with respect to $\zeta_j$, we can easily deduce that the thermal DC conductivity matrix of
the dual CFT is a symmetric matrix.

Fifth, returning to the Stokes equations, if we multiply by
$v^j$ and integrate we get
\begin{align}
\int d^{D-2}x \sqrt{h_{(0)}}2\nabla^iv^jS^{kl}_{ij}\nabla_k v_l=\int d^{D-2}x\delta\tilde Q_H^i\zeta_i\,.
\end{align}
When $\tilde\alpha=0$ the left hand side is positive definite and this is associated with the fact that
the DC thermal conductivity is a positive definite matrix. It would be interesting to examine the behaviour
when $\tilde\alpha\ne 0$.

Sixth, observe that if the horizon admits a Killing vector, then we can solve the source free (i.e. $\zeta=0$) generalised Stokes equations by taking $v^i$ to be the Killing vector, and hence satisfying $\nabla_{(i}v_{j)}=0$, with $p=0$. This means that whenever
the horizon admits a Killing vector then there will not be a unique solution to the Stokes equations.
This is related to the fact that finite DC conductivities require translation invariance to be explicitly broken.

A final observation is that if we define $\bar v^i= (\delta^i_j-4\tilde\alpha {G}^i_{(0)j})v^j$, 
then we have $\nabla_i\bar v^i=0$ and the fluid is incompressible. The Stokes equation 
can be written in terms $\bar v^i$, but as the resulting expression is not particularly illuminating, we omit it.

\subsection{One dimensional lattices}
We now consider the class of background black hole solutions of Gauss-Bonnet gravity in $D$ spacetime dimensions 
that break translations in just one of the spatial directions of the dual field theory.
We assume that the $(D-2)$-dimensional horizon geometry depends on the spatial coordinate $x$ and is independent of the remaining $D-3$ spatial coordinates $x^I$. The horizon metric is given by
\begin{align}\label{onedform}
ds^{2}_{H}=\gamma(x) \,dx^{2}+k_{IJ}(x)dx^I dx^J\,,
\end{align}
and both $k_{IJ}(x)$, $\gamma(x)$ depend periodically on $x$, with period $L$. 
We define $k\equiv \det k_{IJ}$. Note that it is possible to do a coordinate transformation at the horizon
to set $\gamma=1$. However, if we want to use the same spatial coordinate $x$ on the holographic boundary and at the horizon,
which is natural in numerically constructing holographic lattice black holes, generically we have $\gamma\ne 1$.

It is helpful to now define the matrix $M^I{}_J\equiv \gamma^{-1/2}k^{IK}\partial_x k_{KJ}$. We will also raise and lower 
indices via: $M_{IJ}\equiv k_{IK}M^K{}_J$ and $M^{IJ}\equiv k^{IK}M^J{}_K$.  The non-vanishing Christoffel symbols for the horizon metric are given by
\begin{align}
\Gamma^x_{xx}=\frac{1}{2}\partial_x \ln \gamma\,,\quad
\Gamma^x_{IJ}=-\frac{1}{2}\gamma^{-1/2}M_{IJ}\,,\quad
\Gamma^I_{xJ}=\frac{1}{2}\gamma^{1/2}M^{I}{}_{J}\,.
\end{align}
The associated components of the Riemann tensor are given by
\begin{align}
R_{IJKL}&=-\frac{1}{4}(M_{IK}M_{JL}-M_{IL}M_{JK})\,,\nn
R_{xIxJ}&=-\frac{1}{2}\gamma^{1/2}k_{IK}\partial_x{M^K}_J-\frac{1}{4}\gamma M_{IK}{M^K}_J\,,
\end{align}
and $R_{xIJK}=0$.

To proceed we write the perturbed heat current at the horizon as
\begin{align}
\delta \tilde Q^i_H=(4\pi T)\sqrt{h_{(0)}}A^i_j v^j,\qquad A^i_j\equiv \delta^i_j-4\tilde\alpha {G}_{(0)}{}^i_j\,.
\end{align}
We find that $A^x_I=A^I_x=0$ as well as
\begin{align}
A\equiv A^x_x=1+\frac{\tilde{\alpha}}{2}\left[Tr (M^2)-(Tr M)^2\right]\,,
\end{align}
where we have defined $Tr M\equiv {M^K}_K$, $Tr (M^2)\equiv {M^I}_K{M^K}_I$ and so on. 
The condition $\partial_i \delta \tilde Q^i_H=0$ can then be solved with $v^I=0$ and 
\begin{align}
v^x=(\gamma k)^{-1/2}A^{-1}v_0,\qquad \delta\tilde Q^x_H=4\pi T v_0\,,
\end{align}
where $v_0$ is constant.

Next we write the Stokes equations \eqref{nsspecial} as
\begin{align}
-2\nabla_i(B^i_j)= A_j^i[(4\pi T)\zeta_i-\nabla_i p]\,,\qquad B_{ij}\equiv  S_{ij}^{kl}\nabla_k v_l\,.
\end{align}
We now take the $x$ component of this equation, multiply by $A^{-1}$ and then integrate over
a period of $x$. After an
integration by parts we then get
\begin{align}\label{zetaeq2}
4\pi T\zeta_x=\int \,\left[ 2 k^{1/2}B^x_x      \partial_x\left(A^{-1}k^{-1/2}\right)
+A^{-1} B^{IJ}\partial_x k_{IJ}\right]\,.
\end{align}
where $\int\equiv(L)^{-1}\int_0^{L} dx$.
We next notice that we can express the components of $B$ in terms of the constant $v_0$ using the expressions:
\begin{align}
\nabla_x v_x&=(\partial_x-\Gamma^x_{xx})\gamma^{1/2}(g_{d-1})^{-1/2}A^{-1}v_0\,,\nn
\nabla_x v_I&=\nabla_I v_x=0\,,\nn
\nabla_I v_J&=-\Gamma^x_{IJ} \gamma^{1/2}(g_{d-1})^{-1/2}A^{-1}v_0\,.
\end{align}
Thus, \eqref{zetaeq2} can be used to obtain an expression relating $\zeta_x$ in terms of $v_0$, 
and hence in terms of $\delta\tilde Q^x_H$. Since the $\delta\tilde Q^x_H$ is a constant for these one-dimensional lattices, and moreover the bulk currents are independent of
the radius, we can obtain $\delta\tilde Q^x_\infty$ in terms of $\zeta_x$ and hence the thermal conductivity. 

After some lengthy but straightforward calculation we find that at leading order in $\tilde\alpha$ the conductivity can be written as
\begin{align}
\kappa=\frac{(4\pi)^2 T}{X}\,,
\end{align}
where
\begin{align}
X=\int\frac{1}{2}\gamma^{1/2}k^{-1/2}\left[ Tr(M^2)+(Tr M)^2+\tilde\alpha C\right]\,,
\end{align}
with
\begin{align}
C=&-2(D-4)(D-1)[Tr(M^2)+(Tr M)^2)] 
-\frac{2}{3}[Tr M][ Tr (M^3)]\nn
&- Tr(M^4)+\frac{2}{3}(Tr M)^4
+(Tr M)^2 Tr(M^2)\,.
\end{align}
A few comments are in order. Firstly, the final expression is invariant under reparametrisations of the coordinate $x$,
as expected. Second, when $\tilde\alpha=0$ the result is consistent with that derived in section 4.2 of \cite{Banks:2015wha}. Third, when $D=4$, the matrix $M^I{}_J$ is just a number and it is simple to show that $\kappa$ is independent of 
$\tilde\alpha$ which is consistent with the fact that the Gauss-Bonnet
term is topological for $D=4$ and hence does not contribute to the equations of motion. 

In the special case that $k_{IJ}=\phi\delta_{IJ}$, for a periodic function $\phi(x)$, we can simplify the expressions
a little. We find
\begin{align}
X=\int \gamma^{1/2}\frac{(D-3)}{2}\phi^{-\frac{D-3}{2}}\left(\gamma^{-1/2}\partial_x \ln \phi\right)^2
\left(
D-2+\tilde\alpha C\right)\,,
\end{align}
with
\begin{align}
C=(D-4)\left( -2(D-1)(D-2)+\frac{(D-2)(2D-3)}{3}\left(\gamma^{-1/2}\partial_x \ln \phi\right)^2\right)\,.
\end{align}
In order to determine the effect of the Gauss-Bonnet term on the conductivity
one needs to take into account that the horizon metric functions
$\gamma,\phi$ will also receive corrections of order $\tilde\alpha$. It would
certainly be interesting to explore this further.

\section{Final Comments}\label{fincom}

In this paper we have successfully generalised many of the results of \cite{Donos:2015gia,Banks:2015wha,Donos:2015bxe}
concerning DC conductivities to theories of gravity with higher derivative terms. The first main result is the identification of
suitable bulk quantities that allow one to relate current fluxes at the black hole horizon to suitably defined transport current fluxes
of the dual field theory. In section \ref{gendisc} and \ref{gbonsec} we achieved this using a Kaluza-Klein reduction on the time direction. 
The original approach that was used in \cite{Donos:2015gia,Banks:2015wha,Donos:2015bxe}, which came from \cite{Donos:2014cya}, 
worked in a gauge in which the DC sources arise as time dependent perturbations and used a two-form that arises in the derivation of Smarr formula. It would be interesting to see how that approach generalises to higher derivative gravity; for the case of Lovelock gravity the work of \cite{Kastor:2008xb,Kastor:2010gq,Liberati:2015xcp} should be helpful. We also anticipate connections with the work of \cite{Wald:1993nt,Iyer:1994ys}.

The second main result was to obtain a generalised set of Stokes equations on the black hole horizon for the case of
static black hole backgrounds in the context of Gauss-Bonnet gravity. These equations form a closed
set of equations for a subset of the perturbation at the horizon and by solving them one can obtain the thermal currents at the horizon.
It should be possible to generalise these results from the case of static black holes to stationary black holes and it would
be interesting to identify the additional terms that will enter the Stokes equations, including Coriolis terms. 

We obtained the Stokes equations for Gauss-Bonnet theory using a radial Hamiltonian formalism and
it should be straightforward to generalise this to arbitrary Lovelock theories.
For general higher dimensional
theories of gravity one will not have such a formalism to hand. Nevertheless, if we write the Einstein equations as $E_{\mu\nu}=0$ then
by considering the projections $n^\mu n^\nu E_{\mu\nu}=0$ and $n^\mu\sigma_\nu^\rho E_{\mu\rho}=0$, where $n^\mu$ is the normal and $\sigma_\nu^\rho$ are the normal and projector for the radial slicing, and then evaluating at the horizon
should give a generalised closed set of Stokes equations for more general theories.

With the results of this paper it would also be interesting to construct and study specific examples of holographic lattices, to investigate the impact of the higher derivative
couplings on the DC conductivity. 
We note that a specific holographic model involving Gauss-Bonnet gravity coupled to a gauge-field
and massless scalars was investigated in \cite{Cheng:2014tya}. In this work the momentum dissipation
just arises from the massless scalar fields, as in the construction \cite{Andrade:2013gsa}, and it was shown that
the DC thermoelectric conductivity is independent of the Gauss-Bonnet coupling. However, this behaviour is an exception, arising
from the simple mechanism for momentum dissipation.
In a parallel direction, it would also be interesting if our results could be used to place bounds on thermal conductivities along
the lines of \cite{Grozdanov:2015qia,Grozdanov:2015djs,Fadafan:2016gmx}.

\section*{Acknowledgements}
The work is supported by the European Research Council under the European Union's Seventh Framework Programme (FP7/2007-2013), ERC Grant agreement ADG 339140. The work of 
JPG is also supported by STFC grant ST/L00044X/1, EPSRC grant EP/K034456/1,
JPG is also supported as a KIAS Scholar and as a Visiting Fellow at the Perimeter Institute.

\appendix

\section{Currents in the dual CFT}\label{currcomment}
In this appendix we discuss how the currents defined in section \ref{secbhgih}
are related to the dual CFT, including a discussion of the counter-term contributions.

We start with the bulk contributions to the currents. 
To simplify the presentation, we just discuss theories of bulk gravity without higher derivative terms;
the generalisation to theories with higher derivatives is straightforward, using the analysis and discussion at the beginning  of section \ref{gbonsec}.
We recall
the derivation of the two co-closed two forms as given in \eqref{formeqmotion}. In varying the dimensionally reduced 
action \eqref{eq:lower_D_functional} 
we obtain a boundary term and on shell we find
\begin{align}
\delta S=2\int_{M_{D-1}}\,d^{D-1}y\,\sqrt{\gamma_{D-1}}\nabla_m(V^{mn}\delta\beta_n+W^{mn}\delta\alpha_n)\,.
\end{align}
If we define variations at the AdS boundary at $r\to\infty$ via: $\delta\beta_i^\infty=\lim_{r\to\infty}\delta \beta_i$ and
$\delta\alpha_i^\infty=\lim_{r\to\infty}\delta \alpha_i$ we have
\begin{align}\label{blksd}
\frac{\delta S}{\delta\beta_i^\infty}=J_{\infty}^i\,,\qquad
\frac{\delta S}{\delta\alpha_i^\infty}=-Q_{\infty}^i\,,
\end{align}
where $J_{\infty}^{i}$ and $Q^{i}_{\infty}$ as the boundary limits
of the bulk currents defined in \eqref{defjayque}.

Now the full bulk action that we should be considering in holography needs to be supplemented by a boundary contribution:
$S_{Tot}\equiv S+S_{Bdy}$, where $S_{Bdy}$ has two key features. The first is that it has counterterms to ensure that all divergences are cancelled. The second is
that it includes a Gibbons-Hawking term to ensure that
on-shell, $S_{Tot}$ is a functional of the boundary metric, $g_{ab}^\infty$, and gauge-field $A_a^\infty$, where
we have introduced the field theory coordinates, $x^a$, with
\begin{align}
x^a=(t,x^i)\,.
\end{align}
As usual, the holographic stress tensor density of the dual CFT is defined as $[T_{Tot}]^{ab}=2\delta S_{Tot}/\delta g_{ab}^\infty$ and 
the holographic boundary current density as $J^{a}_{Tot}=\delta S_{Tot}/\delta A_{a}^\infty$. Using the chain rule, and recalling
the definition of $\alpha_i$ and $\beta_i$ in \eqref{eq:diff_metric_gfield_ansatz} we have
\begin{align}
\frac{\delta S_{Tot}}{\delta\beta_i^\infty}=J_{Tot}^i\,,\qquad
\frac{\delta S_{Tot}}{\delta\alpha_i^\infty}=-Q_{Tot}^i\equiv ([T_{Tot}]^{i}{}_t+A_tJ^i_{Tot})\,.
\end{align}

We thus see that $J_{\infty}^{i}$, $Q^{i}_{\infty}$ are the contributions from the bulk action to the finite renormalised currents
of the dual CFT given by $J_{Tot}^{i}$, $Q^{i}_{Tot}$. In general $J_{\infty}^{i}$, $Q^{i}_{\infty}$
are divergent quantities. For the background geometry currents given in \eqref{twee}, the contributions from $S_{Bdy}$ will ensure that the total 
magnetisation densities are finite. $S_{Bdy}$ will also give regulating contributions to
the currents that depend on the DC perturbation as given in \eqref{emcasep}. For the specific case of Einstein-Maxwell theory in $D=4,5$ spacetime dimensions,
it was explicitly shown in \cite{Donos:2015bxe} that the contributions from $S_{Bdy}$ regulate the magnetisation terms given in \eqref{deftcs} as well as regulating
the magnetisation currents given on the right hand side of \eqref{lastcexp}. The net effect of the contribution of $S_{Bdy}$ is thus to ensure that
in the expressions for the perturbed current densities and the perturbed transport current densities given in \eqref{deftcs}-\eqref{result} we can replace all boundary quantities with the finite renormalised quantities. 
In particular, associated with \eqref{result} we have that the renormalised transport current flux densities of the dual CFT are the same as the 
horizon current flux densities (which are, of course, finite quantities).
Why this should be true in general is expanded upon below.

Before doing that we calculate the holographic charge densities using the dimensional reduced
formalism of section 2. Starting with the dimensionally reduced action \eqref{eq:lower_D_functional} with \eqref{ellform}, by considering
the on-shell variation of the action with respect to $A_t^\infty$ we find the charge density
\begin{align}
J^t_\infty&=\left[(\gamma^{(B)}_{D-1})^{1/2}\frac{\delta \mathcal{L}}{\delta u_r}\right]_\infty\,.
\end{align}
Similarly, to get $T^{tt}_\infty$ we can vary with respect to $g_{tt}^\infty$. Using the fact that $g_{tt}=-H^2$ 
we find that on-shell
\begin{align}
[T^t{}_t]_\infty=\left[(\gamma^{(B)}_{D-1})^{1/2}H\frac{\delta \mathcal{L}}{\delta h_r}\right]_\infty\,.
\end{align}
With these formulae in hand, we can now work out how they depend on the DC perturbation.
Using \eqref{expsII} and \eqref{eq:forms_pert} 
we immediately deduce that $\delta J^t_\infty$ is independent of $\phi_T$ and $\phi_E$ and
hence is globally defined on a fundamental domain.
On the other hand, with $Q^t_\infty\equiv -\left[T^t_t+A_tJ^t\right]_\infty$, we find
$\delta Q^t_\infty$ is not globally defined but rather that $\delta Q^t_\infty +\phi_TQ^t_{(B)}+\phi_E J^t_{(B)}$ is
globally defined.

\subsection{Counterterm corrections to magnetisation}\label{cterms}
We now show that the counterterm contribution to the boundary action
only gives corrections to the magnetisation terms that
were introduced in section \ref{gendisc}. The argument follows from the existence of a time-like Killing vector at the boundary, defined as the $r\rightarrow \infty$ limit of the bulk Killing vector. If
$x^a=\{t, x^i\}$ are the boundary coordinates we can carry out a Kaluza-Klein reduction on the boundary with respect to 
$\partial_t$.
Using a similar argument as in section \ref{gendisc}, we take the boundary counterterm action to have the following form
\begin{align}
S_{ct}  =\int d^{D-2}x \sqrt{\tilde{\gamma}} \tilde{H} \mathcal{L}_{ct}(\tilde{h}, \tilde{u}, \tilde{v}, \tilde{w}, \tilde{\gamma}_{ij})\,,
\end{align}
where $D$ is the number of bulk dimensions, the metric and gauge field on the boundary are
\begin{align}
d\tilde{s}^2 &= -\tilde{H}^2(dt+ \tilde{\alpha})^2 +\tilde{\gamma}_{ij}dx^i dx^j\,,\nn
\tilde{A}&=\tilde{A}_{t}\,\left( dt+\tilde{\alpha} \right)+\tilde{\beta}\,,
\end{align}
and we have defined
\begin{align}
\tilde{h} = d \log \tilde{H}\,, \qquad
\tilde{u}=\tilde{H}^{-1}d\tilde{A}_{t}, \qquad
\tilde{v}=d\tilde{\beta}+\tilde{A}_{t}\,d\tilde{\alpha},\qquad
 \tilde{w} = \tilde{H} d\tilde{\alpha}\,,
\end{align}
where $\tilde{h},\tilde{u}$ are one-forms and $\tilde{v},\tilde{w}$ are two-forms, all living in $M_{D-2}$, obtained by taking the $r\to\infty$ limit of the bulk quantities. We have used a tilde to distinguish fields defined at the boundary with the bulk fields as used in the draft. 

Now $S_{ct}$ will contribute to the electric and heat current densities via
\begin{align}
J^i_{ct}&
= \frac{\delta S_{ct}}{\delta\tilde{ \beta}_i} =   \partial_j {M}^{ij}_{ct} \,,\nn
Q^i_{ct}&
= -\frac{\delta S_{ct}}{\delta\tilde{ \alpha}_i} =  \partial_j{M}_{Tct}^{ij} \,,
\end{align}
where
\begin{align}
{M}^{ij}_{ct}&
= 2\sqrt{\tilde{\gamma}}\tilde{H}\frac{\delta \mathcal{L}_{ct}}{\delta \tilde{v}_{ij}}\,,\nn
{M}_{Tct}^{ij} &
=-2\sqrt{\tilde{\gamma}}\tilde{H}^2 \frac{\delta \mathcal{L}_{ct}}{\delta \tilde{w}_{ij}}-2\sqrt{\tilde{\gamma}}\tilde{H}\tilde{A}_t \frac{\delta \mathcal{L}_{ct}}{\delta \tilde{v}_{ij}}\,.
\end{align}
Note that $\tilde\alpha_i$, $\tilde\beta_i$ were denoted $\alpha_i^\infty$, $\beta_i^\infty$, respectively,
in \eqref{blksd}.

We now consider the bulk DC perturbation of \eqref{expsII}, where for simplicity we again assume that 
the locally defined functions $\phi_E$ and $\phi_T$ are independent of the radial coordinate.
On the boundary the DC perturbation takes the form 
\begin{align}
&\tilde{H}=\tilde{H}^{(B)}\left(1-{\phi}_{T}\right) +\delta \tilde{H}, \quad \tilde{\alpha}=\tilde{\alpha}^{(B)} \left(1+{\phi}_{T} \right)+\delta \tilde{\alpha}\,,
\quad \tilde{\gamma}_{ij}=\tilde{\gamma}_{ij}^{(B)}+\delta \tilde{\gamma}_{ij}\,,\notag\\
&\tilde{A}_{t}=\tilde{A}_{t}^{(B)}\,\left(1-{\phi}_{T} \right)+{\phi}_{E}+\delta \tilde{A}_{t},\qquad
\tilde{\beta}=\tilde{\beta}^{(B)}-{\phi}_{E}\,\tilde{\alpha}^{(B)}+\delta\tilde{\beta}\,,
\end{align}
with $\delta \tilde{H},\delta\tilde{\alpha},\delta \tilde{A}_t,\delta\tilde{\beta}$ and $\delta \tilde{\gamma}_{ij}$ all globally defined perturbations on $M_{D-2}$. 
Note that $\tilde{h}$, $\tilde{u}$, $\tilde{v}$ and $\tilde{w}$ are all globally defined.
Therefore, for this perturbation we have:
\begin{align}
\delta {M}^{ij}_{ct} &= \bar\delta{M}^{ij}_{ct} - {{M}^{ij}_{ct}}^{(B)}{\phi}_T\,,\nn
\delta{M}_{Tct}^{ij} &= \bar\delta{M}_{Tct}^{ij}- 2 {{M}_{Tct}^{ij}}^{(B)}{\phi}_T-{{M}_{ct}^{ij}}^{(B)}{\phi}_E \,,
\end{align}
where $ \bar\delta{M}^{ij}_{ct}$ and $  \bar\delta{M}_{Tct}^{ij}$ are the pieces independent of $\phi_E,\phi_T$ and hence
are globally-defined densities defined and $ {{M}_{ct}^{ij}}^{(B)}$ and $ {{M}_{Tct}^{ij}}^{(B)}$ are the corresponding values in the background.
We thus see that the counter terms will give corrections that renormalise the magnetisations given in
\eqref{magdefs}. In turn this means, in effect, that in \eqref{deftcs}-\eqref{result} we can replace all boundary quantities with the finite renormalised quantities. 

The above analysis is valid for two derivative theories in the bulk with counterterms that also have two-derivative terms.
If we relax these conditions, either for a higher derivative bulk theory, or for a two derivative theory of gravity
with higher derivative boundary counterterms, then we should carry out suitable generalisations analogous to
the discussion at the beginning  of section \ref{gbonsec}.

To conclude this appendix, we briefly illustrate the above for a counterterm action that appears for theories of gravity with
two derivatives in $D=4,5$:
\begin{align}
S_{ct} = -\frac{1}{D-3}\int dt d^{D-2}x \sqrt{-\tilde\sigma} \left(\tilde R_{D-1} + 2 (D-3)(D-2)\right)\,,
\end{align}
where $\tilde\sigma$ is the $(D-1)$-dimensional boundary metric.
Using the KK decomposition, this leads to 
\begin{align}
 \mathcal{L}_{ct}=-\frac{1}{D-3}\left(R_{D-2} +\frac{1}{4}\tilde{w}^2+ 2 (D-3)(D-2)\right)\,.
\end{align}
From here we obtain 
\begin{align}
\mathcal{M}_{Tct}^{ij} &= \frac{\sqrt{\tilde{\gamma}}\tilde{H}^2 }{(D-3)} w^{ij}\,,
\end{align}
giving a contribution to the current of the form
\begin{align}
Q^i_{ct}&= \partial_j\left(\frac{\sqrt{\tilde{\gamma}}\tilde{H}^2 }{(D-3)} w^{ij}\right)  \,.
\end{align}
This agrees with the result in appendix B of \cite{Donos:2015bxe} (to make the comparison one can use the result
\eqref{ricexps} to write $R^{\hat m}{}_{0}=\tfrac{1}{2}H^{-2}\nabla_{\hat n}(H^2 w^{\hat n \hat m})$).

\section{Currents at the horizon for higher derivative theories}\label{details}
In this appendix we outline a few more details on how we calculated the currents for the
the higher derivative theory of gravity discussed in section \ref{gbonsec}.
For simplicity, as in the text, we consider static backgrounds and set
$\alpha^{(B)}=0$ in \eqref{eq:diff_metric_gfield_ansatz}.

For the background black hole geometries we have the expansions:
\begin{align}\label{nhorexpaapp}
\gamma_{rr}^{(B)}&=\frac{1}{4\pi T\,r}\,\left( 1+\gamma_{rr}^{(1)}\,r\right)+\mathcal{O}(r)\,,\quad
\gamma^{(B)}_{ij}=h^{(0)}_{ij}+h^{(1)}_{ij}\,r+\mathcal{O}(r^{2})\,,\quad
\gamma^{(B)}_{ir}=\mathcal{O}(r)\,,\nn
H^{(B)}{}^{2}&= 4\pi T\,r\,\left(1+2\,H^{(1)}\,r \right)+\mathcal{O}(r^{3})\,.
\end{align}
We thus have 
\begin{align}\label{nhorexpaup}
&\gamma^{(B)rr}={4\pi Tr}\,\left( 1-\gamma_{rr}^{(1)}\,r\right)+\mathcal{O}(r^3),\quad
\gamma^{(B)ri}=\mathcal{O}(r^2),\quad
\gamma^{(B)ij}=h^{(0)ij}-h^{(1)ij}\,r+\mathcal{O}(r^{2}),\quad\nn
&(\gamma^{(B)}_{D-1})^{1/2}=\frac{\sqrt{h^{(0)}}}{(4\pi T r)^{1/2}}\left(1+r\left(\frac{1}{2}{\gamma^{(1)}_{rr}}+\frac{1}{2}h^{(1)}{}^{i}{}_{i}   \right)   \right)+\mathcal{O}(r^{3/2})\,,\nn
&h_r=\frac{1}{2r}+H^{(1)}+\mathcal{O}(r)\,,
\end{align}
where indices on $h^{(1)}$ are raised with $h^{(0)}$ e.g $h^{(1)}{}^{i}{}_{i}=h^{(0)ij}h^{(1)}_{ij}$.
We next expand the Christoffel connection for the $(D-1)$-dimensional metric $\gamma_{D-1}$
at the horizon to find
\begin{align}
\bar\Gamma^{r}_{rr}&=-\frac{1}{2r}+\frac{1}{2}\gamma^{(1)}_{rr}+\mathcal{O}(r)\,,\nn
\bar\Gamma^{r}_{ri}&=\mathcal{O}(r)\,,\nn
\bar\Gamma^{r}_{jk}&=-(4\pi T)\frac{r}{2}h^{(1)}_{jk}+\mathcal{O}(r^2)\,,\nn
\bar\Gamma^{i}_{rr}&=\mathcal{O}(1)\,,\nn
\bar\Gamma^{i}_{rj}&=\frac{1}{2}h^{(1)}{}^i{}_j+\mathcal{O}(r)\,,\nn
\bar\Gamma^{i}_{jk}&=\gamma^{i}_{jk}+\mathcal{O}(r)\,,
\end{align}
where $\gamma^{i}_{jk}$ is associated with $h^{(0)}_{ij}$.
Similarly, the components of the Riemann tensor for the $(D-1)$-dimensional metric $\gamma_{D-1}$
at the horizon are given by
\begin{align}
\bar R^r{}_{rri}&=\mathcal{O}(1)\,,\nn
\bar R^r{}_{rij}&=\mathcal{O}(r)\,,\nn
\bar R^r{}_{irj}&=-\frac{1}{4}(4\pi T) h^{(1)}_{ij}+\mathcal{O}(r)\,,\nn
\bar R^r{}_{kij}&=\mathcal{O}(r)\,,\nn
\bar R^k{}_{rri}&=\frac{1}{4r}h^{(1)}{}^k{}_i+\mathcal{O}(1)\,,\nn
\bar R^k{}_{rij}&=\mathcal{O}(1)\,,\nn
\bar R^k{}_{lri}&=\mathcal{O}(1)\,,\nn
\bar R^k{}_{lij}&=R^{(0)k}{}_{lij}+\mathcal{O}(r)\,.
\end{align}
Hence for the Ricci tensor we have
\begin{align}\label{tink1}
\bar R_{rr}&=-\frac{1}{4r}h^{(1)}{}^i{}_i+\mathcal{O}(1)\,,\nn
\bar R_{ri}&=\mathcal{O}(1)\,,\nn
\bar R_{ij}&=R^{(0)}_{ij}-\frac{1}{4}(4\pi T) h^{(1)}_{ij}+\mathcal{O}(r)\,,
\end{align}
and for the Ricci scalar
\begin{align}\label{tink2}
\bar R&=R^{(0)}-\frac{1}{2}(4\pi T)h^{(1)}{}^i{}_i+\mathcal{O}(r)\,.
\end{align}

Since we are in the static case with $\alpha^{(B)}=0$ we have from \eqref{eq:forms_pert} that
\begin{align}
\delta w&=H^{(B)}\,d\delta\alpha\,,
\end{align}
and hence using \eqref{eq:nh_pert},\eqref{nhorexpaapp},\eqref{nhorexpaup}
we deduce that as $r\to 0$
\begin{align}
\delta w_{ri}&=-\frac{1}{(4\pi T)^{1/2} r^{3/2}} v_i+\frac{1}{(4\pi T)^{1/2} r^{1/2}}\left( \zeta_{i}+\partial_{i}\delta g_{tr}^{(0)}-H^{(1)}v_{i}\right)+\mathcal{O}(r^{1/2})\,,\nn
\delta w_{ij}&=\frac{1}{(4\pi T)^{1/2} r^{1/2}}\,dv_{ij}+\mathcal{O}(r^{1/2})\,,
\end{align}
as well as
\begin{align}\label{wuppa}
\delta w^{ri}&=-(4\pi T)^{1/2}\frac{1}{r^{1/2}}\left(v^i-r\left( \zeta^{i}+\nabla^{i} \delta g_{tr}^{(0)} 
-H^{(1)}v^i+\gamma^{(1)}_{rr}v^i+h^{(1)ij}v_j\right)\right) +\mathcal{O}(r^{3/2})\,,\nn
\delta w^{ij}&=\frac{1}{(4\pi T)^{1/2}}\frac{1}{r^{1/2}}2\nabla^{[i}v^{j]}+\mathcal{O}(r)^{1/2}\,.
\end{align}

As explained in the text, the currents at the horizon are given by
\begin{align}\label{timnow}
\delta Q^i_H&=(4\pi T)\sqrt{h^{(0)}} v^i
+c_1\delta Q^{(1)}_H{}^{i}+c_2\delta Q^{(2)}_H{}^{i}+c_3\delta Q^{(3)}_H{}^{i}\,,
\end{align}
where
\begin{align}
\frac{1}{4\pi T \sqrt{h_{(0)}}}\delta Q^{(a)}_H{}^{i}\equiv& 
-\frac{2r^{1/2}}{(4\pi T)^{1/2}}\,\left[\delta(\delta_{w}\mathcal{L}^{(a)})^{ri}\right]_H\,.
\end{align}
We next calculate each of the three variations appearing on the right hand side of \eqref{timnow}.
We start by analysing $\delta [\delta_{w}\mathcal{L}^{(3)})]^{ri}$ as $r\to 0$.
From \eqref{defgenW}, since $w^{(B)}=0$ in the static background, we find
\begin{align}
\delta[\delta_{w}\mathcal{L}^{(3)}]^{ri}&=R\,\delta w^{ri}\,,\nn
&=(\bar R-2\nabla_m h^m-2h^2)\delta w^{ri}\,,
\end{align}
where we used \eqref{rsexps}.
If we now evaluate at the horizon, we find a cancellation of the $1/r$ singular terms
in $\nabla_m h^m$ and $h^2$. Using \eqref{tink2} as well as \eqref{wuppa}
we obtain
\begin{align}
\frac{1}{4\pi T \sqrt{h_{(0)}}}\delta Q^{(3)}_H{}^{i}
&=R^{(0)}v^i-\,(4\pi T)
\left(3H^{(1)}-\frac{1}{2}\gamma^{(1)}_{rr}+h^{(1)}{}^j{}_j\right)v^{i}\,,
\end{align}
which is the result given in \eqref{qthreefe}.

Next we want to consider the limit of $\delta [\delta_{w}\mathcal{L}^{(2)})]^{ri}$ as $r\to 0$. From the definition 
in \eqref{defgenW} and for a static background with $w^{(B)}=\Lambda^{(1)(B)}=0$, we have
\begin{align}\label{lambtwovar}
\delta [\delta_{w}\mathcal{L}^{(2)}]^{ri}&=-2\,\Lambda^{(2)}{}_{p}{}^{[r}\delta w^{i]p}+H\,d\left(H^{-1}\,\delta \Lambda^{(1)} \right)^{ri}+\Lambda^{(0)}\,\delta w^{ri}\,.
\end{align}
Next consider each of the three terms in \eqref{lambtwovar} as $r\to 0$. We have
\begin{align}\label{pthree}
-2\,\Lambda^{(2)}{}_{p}{}^{[r}\delta w^{i]p}&=\Lambda^{(2)}{}_r{}^r\delta w^{ri}
+\Lambda^{(2)}{}_j{}^i\delta w^{rj}+\Lambda^{(2)}{}_j{}^r\delta w^{ij}\,,\nn
&\to -(4\pi T)\frac{1}{2}\left(\frac{1}{2}h^{(1)}{}_i{}^i+3 H^{(1)}-\frac{1}{2}\,\gamma_{rr}^{(1)} \right)\delta w^{ri}
+(R^{(0)}_j{}^i-\frac{(4\pi T)}{2}h^{(1)}_j{}^i)\delta w^{rj}\,,
\end{align}
as $r\to 0$, and we note in particular that the last term in the first line does not contribute.
We next calculate
\begin{align}
\delta\Lambda^{(1)}{}^{i}&=\frac{1}{2\,r^{1/2}}\,\frac{1}{(4\pi T)^{1/2}}\left(2\,\nabla_{j}\nabla^{[j}v^{i]}+4\pi T\left( \zeta^{i}+\nabla^{i} \delta g_{tr}^{(0)}\right)-M^{ij}v_{j} \right)+\mathcal{O}(r^{1/2})\,,\nn
\delta\Lambda^{(1)}{}^{r}&=\frac{(4\pi T)^{1/2}}{2\,r^{1/2}}\,\nabla_{i}v^{i}+\mathcal{O}(r^{1/2})\,,
\end{align}
where, as in appendix D of \cite{Banks:2015wha},
\begin{align}
M^{ij}=(4\pi T)\left(h^{(0)}{}^{ij}\,\left(3\,H^{(1)}-\frac{1}{2}\,\gamma_{rr}^{(1)} \right)+\frac{1}{2}h^{(0)ij}h^{(1)}{}^k{}_k 
-h^{(1)ij}\right)\,.
\end{align}
We then find that we can write, as $r\to 0$, 
\begin{align}\label{ptwo}
H\,d\left(H^{-1}\delta\Lambda^{(1)} \right)^{ri}&\to -(4\pi T)\delta\Lambda^{(1)i}-\nabla^i\delta\Lambda^{(1)r}\,.
\end{align}
As $r\to 0$, we also have
\begin{align}\label{pone}
\Lambda^{(0)}\,\delta w^{ri}&\to \frac{(4\pi T)}{2}\left[3\,H^{(1)}-\frac{1}{2}\,\gamma_{rr}^{(1)} +
\frac{1}{2}h^{(1)}{}^k{}_k\right]\delta w^{ri}\,,
\end{align}
which will cancel a term in \eqref{lambtwovar}.
Combining all of these three terms we conclude that as $r\to 0$ we have 
\begin{align}
&\delta [\delta_{w}\mathcal{L}^{(2)}]^{ri}=-\frac{(4\pi T)^{1/2}}{r^{1/2}}\,R_{(0)}^{ij}v_{j}+\frac{(4\pi T)^{3/2}}{r^{1/2}}\frac{1}{2}h^{(1)}{}^{ij}v_j-\frac{(4\pi T)^{1/2}}{r^{1/2}}\,\frac{1}{2}\,\nabla^{i}\nabla_{j}v^{j}\notag\\
&-\frac{(4\pi T)^{1/2}}{r^{1/2}}\,\frac{1}{2}\left(2\,\nabla_{j}\nabla^{[j}v^{i]}+4\pi T\left( \zeta^{i}+\nabla^{i} \delta g_{tr}^{(0)}\right)-M^{ij}v_{j} \right)
+\mathcal{O}(r^{1/2})\,.
\end{align}
This leads to the expression for $\delta Q_H^{(2)}{}^{i}$ given in \eqref{qtwofe}.

Finally, we want to calculate the limit of $\delta [\delta_{w}\mathcal{L}^{(1)})]^{ri}$ as $r\to 0$. From the definition 
in \eqref{defgenW} and for a static background with $w^{(B)}=\Sigma^{(3)(B)}=0$, we have 
\begin{align}\label{vlone}
\delta[\delta_{w}\mathcal{L}^{(1)}]^{ri}&=2\left(\Sigma^{(4)}{}^{ri}{}_{pq}+\,\Sigma^{(4)}{}^{\left[r\right.}{}_{p}{}^{\left. i\right]}{}_{q} \right)\,\delta w^{pq}\notag\\
&\quad -4\,h_{p}\,\left(\delta \Sigma^{(3)}{}^{pri}-2\,\delta \Sigma^{(3)}{}^{\left[ ri \right]\,p} \right)+4H^{-1}\,\nabla_{p}\left(H\, \delta \Sigma^{(3)}{}^{pri}\right)\nn
&\quad -4\,\Sigma^{(2)}{}^{p\left[ r\right.}w^{\left. i\right]}{}_{p}\,.
\end{align}
For the first line of \eqref{vlone}, as $r\to 0$ we find
\begin{align}\label{fline}
2\left(\Sigma^{(4)}{}^{ri}{}_{pq}+\,\Sigma^{(4)}{}^{\left[r\right.}{}_{p}{}^{\left. i\right]}{}_{q} \right)\delta w^{pq}
&\to 6 \bar R^{ri}{}_{rj}\delta w^{rj}\,,\nn
&=(4\pi T)^{3/2}\frac{3}{2}\frac{1}{r^{1/2}}h^{(1)ij}v_j+\mathcal{O}(r^{1/2})\,.
\end{align}
Next, the third line in \eqref{vlone} is given by
\begin{align}
-4\Sigma^{(2)}{}_p{}^{[r}\delta w^{i]p}=-2\Sigma^{(2)}{}_j{}^{r}\delta w^{ij}+2\Sigma^{(2)}{}_r{}^{r}\delta w^{ri}
+2\Sigma^{(2)}{}_j{}^{i}\delta w^{rj}\,.
\end{align}
In the background we calculate that as $r\to 0$ we have
\begin{align}
\Sigma^{(2)}_{rr}\to \frac{1}{2r}\left(3 H^{(1)}-\frac{1}{2}\gamma^{(1)}_{rr}\right)\,,\quad
\Sigma^{(2)}_{ij}\to(4\pi T)\frac{1}{4}h^{(1)}_{ij}\,,\quad
\Sigma^{(2)}_{ri}=\mathcal{O}(1)\,,
\end{align}
and hence as $r\to 0$
\begin{align}\label{thline}
-4\Sigma^{(2)}{}_p{}^{[r}\delta w^{i]p}&=
-\frac{(4\pi T)^{3/2}}{r^{1/2}}\left[(3H^{(1)}-\frac{1}{2}\gamma^{(1)}_{rr})v^i+\frac{1}{2}h^{(1)ij}v_j\right]
+\mathcal{O}(r^{1/2})\,.
\end{align}
We next consider the second line in \eqref{vlone} which can be written 
\begin{align}\label{seclll}
 -4\,h_{p}\,&\left(\delta \Sigma^{(3)}{}^{pri}-2\,\delta \Sigma^{(3)}{}^{\left[ ri \right]\,p} \right)+4H^{-1}\,\nabla_{p}\left(H\,\delta \Sigma^{(3)}{}^{pri}\right)
 =\nn
 &-8h_r\delta\Sigma^{(3)rri}+4H^{-1}\,\nabla_r\left(H\,\delta\Sigma^{(3)rri}\right)+4H^{-1}\,\nabla_j\left(H\,\delta\Sigma^{(3)jri}\right)\,.
 \end{align}
To proceed we evaluate the limits 
\begin{align}
\delta\Sigma^{(3)rij}=\mathcal{O}(r^{1/2})\,,\qquad
\delta\Sigma^{(3)ijk}=\mathcal{O}(r^{-1/2})\,,
\end{align}
as well as 
\begin{align}
&
\delta \Sigma^{(3)rri}=r^{1/2}(4\pi T)^{3/2}\frac{1}{2}\left(-(3H^{(1)}-\frac{1}{2}\gamma^{(1)}_{rr})v^i
+\frac{1}{2} h^{(1)ij}v_j+(\zeta^i+\nabla^i\delta g_{tr}^0)\right)+\mathcal{O}(r^{3/2})\,,\nn
&
\delta\Sigma^{(3)jri}
=-(4\pi T)^{1/2}\frac{1}{2}\frac{1}{r^{1/2}}\nabla^{(j} v^{i)}+\mathcal{O}(r^{1/2})\,.
\end{align}
A calculation then reveals as $r\to 0$ 
\begin{align}
-8h_r\delta\Sigma^{(3)rri}+4H^{-1}\,\nabla_r\left(H\,\delta\Sigma^{(3)rri}\right)\to -\frac{4}{r}\delta\Sigma^{(3)rri}\,,
\end{align}
and
\begin{align}
4H^{-1}\nabla_j(H\delta\Sigma^{(3)jri})=4\Bigg(-(4\pi T)^{1/2}\frac{1}{2}\frac{1}{r^{1/2}}\nabla_j\nabla^{(j} v^{i)}\Bigg)+\mathcal{O}(r^{1/2})\,.
\end{align}
After substituting into \eqref{seclll} we obtain the second line of \eqref{vlone}. Finally combining the resulting
expression with \eqref{fline} and \eqref{thline}, we deduce that as $r\to 0$ \eqref{vlone} can be written as
\begin{align}
\delta(\delta_{w}&\mathcal{L}^{(1)})^{ri}=-2\frac{(4\pi T)^{1/2}}{r^{1/2}}\,\nabla_{j}\nabla^{(j}v^{i)}\nn
&-\frac{(4\pi T)^{3/2}}{r^{1/2}}\left[ 2\left(\zeta^{i}+\nabla^{i}\delta g_{tr}^{(0)}\right)
-\left(3\,H^{(1)}-\frac{1}{2}\gamma_{rr}^{(0)}\right) v^{i}\right]\,.
\end{align}
This leads to the expression for $\delta Q^{(3)}_H{}^{i}$ given in \eqref{qonefe}.

\section{Constraints at the horizon for Gauss-Bonnet}\label{conGB}

We write the spacetime coordinates as $y^\mu=(r,x^a)$, where 
\begin{align}
x^a=(t,x^i)\,,
\end{align}
are the coordinates for the dual field theory.
We perform a radial decomposition of the bulk metric in a standard way, writing
\begin{align}
ds^2&=g_{\mu\nu}dy^\mu dy^\nu=N^2dr^2+\s_{ab}(dx^a+N^adr)(dx^b+N^bdr)\,.
\end{align}
The normal vector to surfaces of constant $r$ has components $n^\mu=N^{-1}(1,-N^a)$, while
$n_\mu=N(1,0)$. The induced metric on the surfaces of constant $r$ is given by $\sigma_{\mu\nu}=g_{\mu\nu}-n_\mu n_\nu$ and has non-vanishing components $\sigma_{ab}$.
The extrinsic curvature is defined as $K_{\mu\nu}=\tfrac{1}{2}{\cal L}_n \s_{\mu\nu}$ and
has non-vanishing components given by 
\begin{align}
K_{ab}=\frac{1}{2N}(\partial_r\s_{ab}-D_aN_b-D_bN_a)\,,
\end{align}
where $N_a=\s_{ab}N^b$. 

We are interested in examining the behaviour of the perturbed geometry near the horizon. We will need
to consider the expansions for the background and the perturbation separately.
From \eqref{nhorexpaapp} the background metric $\s_{ab}$ has the following expansion near the horizon:
\begin{align}
\s_{tt}=-U(1+2H^{(1)}r)+\mathcal{O}(r^2)\,,\qquad
\s_{ij}=h_{ij}^{(0)}+ h_{ij}^{(1)}r+\mathcal{O}(r^2)\,,
\end{align}
with $\s_{ti}=0$. In this appendix we are using 
\begin{align}
U=4\pi T r\,,
\end{align}
for convenience.
From \eqref{eq:nh_pert} the perturbed metric $\delta \s_{ab}$ near the horizon behaves as:
\begin{align}
\delta\s_{tt}&=U\delta g_{tt}^{(0)}+o(r^2)\,,\cr
\delta\s_{ij}&=\delta g_{ij}^{(0)}+o(r)\,,\cr
\delta\s_{ti}&=-v_i-r\ln r\zeta_i-Ut\zeta_i+o(r)+\mathcal{O}(r^2)\,.
\end{align}
Here we use $o(r^n)$ to denote time-independent terms of order $r^n$ or higher, while $\mathcal{O}(r^n)$ also possibly includes time-dependent terms.
Near the horizon, for the background from \eqref{nhorexpaapp} we also have 
\begin{align}
N^2=&\frac{1}{U}(1+\gamma_{rr}^{(1)}r)+\mathcal{O}(r)\,,\cr
N_i=&\mathcal{O}(r)\,,\cr
N_t=&0\,,
\end{align}
while for the perturbation, from \eqref{eq:nh_pert} we have
\begin{align}
\delta N=&\frac{1}{2\sqrt{U}}\delta g_{rr}^{(0)}+\mathcal{O}(r^{1/2})\,,\cr
\delta N_i=&-\frac{v_i}{U}+\mathcal{O}(1)\,,\cr
\delta N_t=&\delta g_{tr}^{(0)}+\mathcal{O}(r)\,,
\end{align}
where $\delta g_{tt}^{(0)}+\delta g_{rr}^{(0)}=2\delta g_{tr}^{(0)}$.

Near the horizon, the components of the Christoffel symbols for the perturbed metric $\s_{ab}+\delta\s_{ab}$ on the radial slices are given by
\begin{align}
\Gamma_{tt}^i=&-U(\zeta^i+\frac{1}{2}\hat{\nabla}^i\delta g_{tt}^{(0)})+\mathcal{O}(r^2)\,,\cr
\Gamma_{ti}^t=&-\frac{1}{2}\hat{\nabla}_i\delta g_{tt}^{(0)}+\mathcal{O}(r)\,,\cr
\Gamma_{ij}^k=&\tilde{\Gamma}_{ij}^k+\mathcal{O}(r)\,,\cr
\Gamma_{tt}^t=&-r v^i\partial_iH^{(1)}+\mathcal{O}(r^2)\,,\cr
\Gamma_{ti}^j=&\frac{1}{2}(\hat{\nabla}^jv_i-\hat{\nabla}_iv^j)+o(r)+\mathcal{O}(r^2)\,,\cr
\Gamma_{ij}^t=&\frac{1}{2U}(\hat{\nabla}_iv_j+\hat{\nabla}_jv_i)+\frac{t}{2}(\hat{\nabla}_i\zeta_j+\hat{\nabla}_j\zeta_i)+o(1)+\mathcal{O}(r)\,.
\end{align}
In the above and in what follows, all tilde quantities are defined with respect to the full horizon metric $\tilde{h}_{ij}^{(0)}\equiv h_{ij}^{(0)}+\delta g_{ij}^{(0)}$. All indices on the tilde quantities are raised and lowered using the full horizon metric $\tilde{h}_{ij}^{(0)}\equiv h_{ij}^{(0)}+\delta g_{ij}^{(0)}$ as well. On the other hand hatted quantities refer to the horizon metric $h_{ij}^{(0)}$ and we have also raised all indices without a tilde using this metric too. In the main text we have dropped the hats to simplify the presentation.

The components of the Riemann tensor for the perturbed metric 
$\s_{ab}+\delta\s_{ab}$ are given by\footnote{Recall that $\hat{\nabla}_j\zeta_i=\hat{\nabla}_i\zeta_j$. We have also used the Bianchi identity ${\,\hat{R}^l}_{ijk}+{\,\hat{R}^l}_{jki}+{\,\hat{R}^l}_{kij}=0$.}:
\begin{align}
{\mathcal{R}^i}_{tjt}=&-U\hat{\nabla}_j(\zeta^i+\frac{1}{2}\hat{\nabla}^i\delta g_{tt}^{(0)})+\mathcal{O}(r^2)\,,\cr
{\mathcal{R}^i}_{jkl}=&{\,\tilde{R}^i}_{jkl}+\mathcal{O}(r)\,,\cr
{\mathcal{R}^t}_{itj}=&\hat{\nabla}_j(\zeta_i+\frac{1}{2}\hat{\nabla}_i\delta g_{tt}^{(0)})+\mathcal{O}(r)\,,\cr
{\mathcal{R}^t}_{tti}=&r v^j \hat{\nabla}_i\hat{\nabla}_jH^{(1)}+\mathcal{O}(r^2)\,,\cr
{\mathcal{R}^t}_{tij}=&\mathcal{O}(r)\,,\cr
{\mathcal{R}^t}_{ijk}=&\frac{1}{2U}\hat{\nabla}_i(\hat{\nabla}_jv_k-\hat{\nabla}_kv_j)-\frac{1}{U}{\,\hat{R}^l}_{ijk}v_l+\mathcal{O}(1)\,,\cr
{\mathcal{R}^i}_{tjk}=&\frac{1}{2}\hat{\nabla}^i(\hat{\nabla}_jv_k-\hat{\nabla}_kv_j)+\mathcal{O}(r)\,,\cr
{\mathcal{R}^i}_{jtk}=&-\frac{1}{2}\hat{\nabla}_k(\hat{\nabla}^iv_j-\hat{\nabla}_jv^i)+\mathcal{O}(r)\,.
\end{align}
The components of the Ricci tensor are:
\begin{align}
\mathcal{R}^t_t=&\hat{\nabla}_i(\zeta^i+\frac{1}{2}\hat{\nabla}^i\delta g_{tt}^{(0)})+\mathcal{O}(r)\,,\cr
\mathcal{R}^i_j=&\tilde{R}^i_j+\hat{\nabla}_j(\zeta^i+\frac{1}{2}\hat{\nabla}^i\delta g_{tt}^{(0)})+\mathcal{O}(r)\,,\cr
\mathcal{R}^t_i=&-\frac{1}{U}{v^j\hat{R}_{ij}}-\frac{1}{2U}\hat{\nabla}_j(\hat{\nabla}^jv_i-\hat{\nabla}_iv^j)+\mathcal{O}(1)\,,\cr
\mathcal{R}^i_t=&\frac{1}{2}\hat{\nabla}_j(\hat{\nabla}^jv^i-\hat{\nabla}^iv^j)+\mathcal{O}(r)\,,
\end{align}
and the Ricci scalar is
\begin{align}
\mathcal{R}=\tilde{R}+2\hat{\nabla}_i(\zeta^i+\frac{1}{2}\hat{\nabla}^i\delta g_{tt}^{(0)})+\mathcal{O}(r)\,.
\end{align}

We next calculate the extrinsic curvature components $K^a_b$ and find\footnote{Note that these correct a typo in the last line of (B.4)
in \cite{Banks:2015wha} as well making (B.5) more precise.}
\begin{align}
K^t_t=&\frac{\sqrt{U}}{2r}(1-\frac{1}{2}\delta g_{rr}^{(0)})+\mathcal{O}(r^{1/2})\,,\cr
K^i_j=&\frac{1}{\sqrt{U}}h^{ik}_{(0)}\hat{\nabla}_{(j}v_{k)}+\mathcal{O}(r^{1/2})\,,\cr
K^i_t=&\frac{\sqrt{U}}{2r}v^i+\mathcal{O}(r^{1/2})\,,\cr
K^t_i=&\frac{\sqrt{U}}{2r}t\zeta_i+o(r^{-1/2})+\mathcal{O}(r^{1/2})\,,\cr
K=&\frac{\sqrt{U}}{2r}(1-\frac{1}{2}\delta g_{rr}^{(0)})+\frac{1}{\sqrt{U}}\hat{\nabla}_iv^i+\mathcal{O}(r^{1/2})\,.
\end{align}

\subsection{Calculation of conjugate momentum
}
We now calculate the conjugate momenta on the horizon, which are given by
\begin{align}
{\pi}^b_a=\sqrt{-\s}(K\delta^b_a-K^b_a)+\tilde \alpha{\pi_{GB}}^b_a\,,
\end{align}
where
\begin{align}\label{gbexpkr}
\frac{1}{\sqrt{-\s}}{\pi_{GB}}^b_a=&2K^b_a(K^2-K^c_dK^d_c-\mathcal{R})\cr
&-4K\mathcal{R}^b_a+4K^c_a\mathcal{R}_c^b+4K_c^b\mathcal{R}^c_a +4K^{cd}{\mathcal{R}^b}_{cad} -4KK_c^bK^c_a+4K_c^bK^c_dK^d_a\cr
&+\delta^b_a(2K\mathcal{R}-\frac{2}{3}K^3+2KK_c^dK^c_d-4K_c^d\mathcal{R}^c_d-\frac{4}{3}K_c^dK^c_eK_d^e)\,.
\end{align}
We calculate each component of ${\pi}^b_a$ in a straightforward manner. 
As the calculations are rather long we have recorded a few intermediate steps.

\subsubsection{Calculation of $\pi^i_t$}
By writing out each of the components and using the previous results we find
\begin{align}
2K^i_t(K^2-K_{ab}K^{ab}-\mathcal{R})
=&2K^i_t(2K_t^tK_j^j-\mathcal{R}_k^k)+\mathcal{O}(r^{1/2})\,,
\end{align}
as well as
\begin{align}
-4K\mathcal{R}^i_t+4K^a_t\mathcal{R}_a^i+&4K^i_a\mathcal{R}^a_t +4K^{ab}{\mathcal{R}^i}_{atb} -4KK_a^iK^a_t+4K_a^iK^a_bK^b_t\cr
&\qquad=-4K_j^j(K^i_tK_t^t)+4K_t^j\mathcal{R}^i_j +\mathcal{O}(r^{1/2})\,,
\end{align}
We then find the final result
\begin{align}\label{piit}
\pi^i_t=-\sqrt{h^{(0)}}\frac{U}{r}v^j(\frac{1}{2}\delta^i_j-2\tilde\alpha \hat{G}^i_j)+\mathcal{O}(r)\,,
\end{align}
where $\hat{G}_{ij}=\hat R_{ij}-\tfrac{1}{2}h^{(0)}_{ij}\hat R$ is the Einstein tensor for the horizon metric
$h^{(0)}_{ij}$.

\subsubsection{Calculation of $\pi_i^t$}
We now have
\begin{align}
2K_i^t(K^2-K_{ab}K^{ab}-\mathcal{R})
=&2K_i^t(2K_t^tK_j^j-\mathcal{R}_j^j)+\mathcal{O}( r^{1/2})\,,
\end{align}
and
\begin{align}
-4K\mathcal{R}^t_i+4K^t_a\mathcal{R}^a_i+&4K^a_i\mathcal{R}^t_a +4K^{ab}{\mathcal{R}^t}_{aib} -4KK_i^aK^t_a+4K^a_iK_a^bK_b^t\cr
&\qquad=-4K_k^kK_i^tK^t_t+4K^t_j\mathcal{R}^j_i +o(r^{-1/2})+\mathcal{O}( r^{1/2})\,.
\end{align}
Combining these we get
\begin{align}
{\pi_{GB}}_i^t
=&\sqrt{h^{(0)}}(2\frac{U}{r}t\zeta_j\hat{G}^j_i)+o(1)+\mathcal{O}( r)\,,
\end{align}
and hence find the following result for the time derivative
\begin{align}
\partial_t\pi^t_i=&-\sqrt{h^{(0)}}\frac{U}{r}\zeta_j(\frac{1}{2}\delta_i^j-2\tilde\alpha\hat{G}^j_i)+\mathcal{O}(r)\,.
\end{align}

\subsubsection{Calculation of $\pi_t^t$}
Now we have
\begin{align}
2K^t_t(K^2-K_{ab}K^{ab}-\mathcal{R})
=&2K_i^t(2K_t^tK_j^j-\mathcal{R}_j^j)+\mathcal{O}( r^{1/2})\,,
\end{align}
as well as
\begin{align}
-4K\mathcal{R}^t_t+4K^t_a\mathcal{R}^a_t+4K^a_t\mathcal{R}^t_a +&4K^{ab}{\mathcal{R}^t}_{atb} -4KK_t^aK^t_a+4K^a_tK_a^bK_b^t\cr
&\qquad=4K^t_t\mathcal{R}^t_t-4K_i^iK_t^tK^t_t+\mathcal{O}(r^{1/2})\,,
\end{align}
and
\begin{align}
&2K\mathcal{R}-\frac{2}{3}K^3+2KK_a^bK^a_b-4K_a^b\mathcal{R}^a_b-\frac{4}{3}K_a^bK^a_cK_b^c\cr
&\qquad=2(-K_t^t\mathcal{R}_t^t+K_t^t\mathcal{R}_i^i+K_j^j\mathcal{R}_i^i-2K_i^j\mathcal{R}^i_j)-2K_t^tK_i^iK_j^j+2K_t^tK_i^jK^i_j+\mathcal{O}(r^{1/2})\,.\cr
\end{align}
Combining these we are led to
\begin{align}
\pi_t^t
=&\sqrt{h^{(0)}}\hat{\nabla}_{i}v^j(\delta_j^i-4\tilde\alpha \hat{G}^i_j)+\mathcal{O}(r)\,.
\end{align}

\subsubsection{Calculation of $\pi^i_j$}
We have
\begin{align}
2K^i_j(K^2-K_{ab}K^{ab}-\mathcal{R})=&2K^i_j(2K_k^kK_t^t+K_k^kK_l^l-K_k^lK^k_l-\mathcal{R}_t^t-\mathcal{R}_k^k)\nn
=&\frac{1}{\sqrt{U}}h^{ik}_{(0)}\hat{\nabla}_{(j}v_{k)}({h^l_l}^{(1)}\frac{U}{r}-2\hat{R})+{h^i_j}^{(1)}\frac{1}{\sqrt{U}}\hat{\nabla}_kv^k\frac{U}{r}+\mathcal{O}(r^{1/2})
\end{align}
as well as
\begin{align}
&-4K\mathcal{R}^i_j+4K^a_j\mathcal{R}_a^i+4K_a^i\mathcal{R}^a_j +4K^{ab}{\mathcal{R}^i}_{ajb} -4KK_a^iK^a_j+4K_a^iK^a_bK^b_j\nn
&=-4K_t^t\mathcal{R}^i_j-4K_k^k\mathcal{R}^i_j+4K_j^k\mathcal{R}^i_k +4K^i_k\mathcal{R}_j^k +4K^{tt}{\mathcal{R}^i}_{tjt}+4K^{kl}{\mathcal{R}^i}_{kjl}-4K_t^tK^i_kK_j^k+\mathcal{O}(r^{1/2})\nn
&=-2\frac{\sqrt{U}}{r}\tilde{R}^i_j+\frac{\sqrt{U}}{r}\delta g_{rr}^{(0)}\hat{R}^i_j-4\frac{1}{\sqrt{U}}\hat{\nabla}_kv^k\hat{R}^i_j+4\frac{1}{\sqrt{U}}\hat{\nabla}_{(j}v_{k)}\hat{R}^{ik} \nn
&\qquad +4\frac{1}{\sqrt{U}}\hat{\nabla}^{(k}v^{i)}\hat{R}_{jk}+4\frac{1}{\sqrt{U}}\hat{\nabla}^{(k}v^{l)}{\,\hat{R}^i}_{kjl}-\frac{\sqrt{U}}{r}\hat{\nabla}^{(i}v^{k)}{h_{jk}}^{(1)}-\frac{\sqrt{U}}{r}{h^{ik}}^{(1)}\hat{\nabla}_{(j}v_{k)}+\mathcal{O}(r^{1/2})
\end{align}
and
\begin{align}
&\delta^i_j(2K\mathcal{R}-\frac{2}{3}K^3+2KK_a^bK^a_b-4K_a^b\mathcal{R}^a_b-\frac{4}{3}K_a^bK^a_cK_b^c)\nn
&\qquad=2\delta^i_j(-K_t^t\mathcal{R}_t^t+K_t^t\mathcal{R}_k^k+K_l^l\mathcal{R}_k^k-2K_k^l\mathcal{R}^k_l-K_t^tK_k^kK_l^l+K_t^tK_k^lK^k_l)+\mathcal{O}(r^{1/2})\nn
&\qquad=2\delta^i_j(\frac{\sqrt{U}}{2r}\tilde{R}+\frac{\sqrt{U}}{2r}(-\frac{1}{2}\delta g_{rr}^{(0)})\hat{R}+\frac{1}{\sqrt{U}}\hat{\nabla}_lv^l \hat{R}\cr
&\qquad\qquad-2\frac{1}{\sqrt{U}}\hat{\nabla}_kv_l\hat{R}^{kl}-\frac{\sqrt{U}}{2r}{h_k^k}^{(1)}\hat{\nabla}_lv^l +\frac{\sqrt{U}}{2r}{h^{kl}}^{(1)}\hat{\nabla}_kv_l)+\mathcal{O}(r^{1/2})\,.
\end{align}
Putting these together we find
\begin{align}
\frac{1}{\sqrt{-\s}}{\pi}_j^i=&(\frac{\sqrt{U}}{2r}(1-\frac{1}{2}\delta g_{rr}^{(0)})+\frac{1}{\sqrt{U}}\hat{\nabla}_kv^k)\delta_j^i-\frac{1}{\sqrt{U}}h^{ik}_{(0)}\hat{\nabla}_{(j}v_{k)}+\tilde\alpha\frac{1}{\sqrt{-\s}}{\pi_{GB}}_j^i+\mathcal{O}(r^{1/2})
\end{align}
where
\begin{align}
&\frac{1}{\sqrt{-\s}}{\pi_{GB}}_j^i=\frac{1}{\sqrt{U}}h^{ik}_{(0)}\hat{\nabla}_{(j}v_{k)}{h^l_l}^{(1)}\frac{U}{r}+{h^i_j}^{(1)}\frac{1}{\sqrt{U}}\hat{\nabla}_kv^k\frac{U}{r}\cr
&-2\frac{\sqrt{U}}{r}\tilde{G}^i_j+\frac{\sqrt{U}}{r}\delta g_{rr}^{(0)}\hat{G}^i_j-4\frac{1}{\sqrt{U}}\hat{\nabla}_kv^k\hat{R}^i_j \cr
& +4\frac{1}{\sqrt{U}}\hat{\nabla}_{(j}v_{k)}\hat{G}^{ik}+4\frac{1}{\sqrt{U}}\hat{\nabla}^{(k}v^{i)}\hat{R}_{jk}+4\frac{1}{\sqrt{U}}\hat{\nabla}^{(k}v^{l)}{\,\hat{R}^i}_{kjl}-\frac{\sqrt{U}}{r}\hat{\nabla}^{(i}v^{k)}{h_{jk}}^{(1)}-\frac{\sqrt{U}}{r}{h^{ik}}^{(1)}\hat{\nabla}_{(j}v_{k)}\cr
&+2\delta^i_j(-2\frac{1}{\sqrt{U}}\hat{\nabla}_kv_l\hat{G}^{kl}-\frac{\sqrt{U}}{2r}{h_k^k}^{(1)}\hat{\nabla}_lv^l +\frac{\sqrt{U}}{2r}{h^{kl}}^{(1)}\hat{\nabla}_kv_l)+\mathcal{O}(r^{1/2})\,.
\end{align}

\subsection{Constraints on the horizon}
The bulk momentum constraint equations $H_a=0$, where $H^a$ is given in \eqref{momcon}, can be written
as
\begin{align}
\sqrt{-\s}\partial_a\left(\frac{\pi^a_b}{\sqrt{-\s}}\right)+\Gamma_{ac}^a\pi^c_b-\Gamma_{ab}^c\pi^a_c=&0\,.
\end{align}
We now evaluate these constraints on a surface of constant $r$ near the horizon and then take the limit $r\to 0$.

\subsubsection{Time component of the momentum constraint}

For the time component, $H_t=0$, we find
\begin{align}
\hat{\nabla}_i\pi^i_t+\mathcal{O}(r)=&0\,,
\end{align}
where we recall that hat refers to the metric on the horizon $h_{ij}^{(0)}$.
Using \eqref{piit} we deduce that
\begin{align}\label{incompress}
(\delta^i_j-4\tilde\alpha \hat{{G}}^i_j)\hat{\nabla}_iv^j=&0\,.
\end{align}
We see that when $\tilde\alpha\ne 0$, the simple incompressibility condition for the fluid
is modified. Alternatively, we can define $\bar v^i= (\delta^i_j-4\tilde\alpha \hat{G}^i_{j})v^j$, 
then we have $\nabla_i\bar v^i=0$ and the fluid is incompressible.

\subsubsection{Spatial component of the momentum constraint}
We next consider the spatial component of the momentum constraint, $H_j=0$, and find that near the horizon we have
\begin{align}
\sqrt{-\s}\tilde{\nabla}_i\left(\frac{\pi^i_j}{\sqrt{-\s}}\right)+\partial_t\pi^t_j+\Gamma_{ti}^t\pi^i_j+\mathcal{O}(r)&=0\,.
\end{align}
After some calculation we find
\begin{align}
0=&\hat{\nabla}_i(\hat{\nabla}_kv^k\delta_j^i-h^{ik}_{(0)}\hat{\nabla}_{(j}v_{k)})\cr
&+\tilde\alpha\hat{\nabla}_i\Big[h^{ik}_{(0)}\hat{\nabla}_{(j}v_{k)}{h^l_l}^{(1)}\frac{U}{r}+{h^i_j}^{(1)}\hat{\nabla}_kv^k\frac{U}{r}-4\hat{\nabla}_kv^k\hat{R}^i_j \cr
& +4\hat{\nabla}_{(j}v_{k)}\hat{{G}}^{ik}+4\hat{\nabla}^{(k}v^{i)}\hat{R}_{jk}+4\hat{\nabla}^{(k}v^{l)}{\,\hat{R}^i}_{kjl}-\frac{U}{r}\hat{\nabla}^{(i}v^{k)}{h_{jk}}^{(1)}-\frac{U}{r}{h^{ik}}^{(1)}\hat{\nabla}_{(j}v_{k)}\cr
&+2\delta^i_j(-2\hat{\nabla}_kv^l\hat{{G}}^k_l-\frac{U}{2r}{h_k^k}^{(1)}\hat{\nabla}_lv^l +\frac{U}{2r}{h^{kl}}^{(1)}\hat{\nabla}_kv_l)\Big]\cr
&-\frac{U}{r}\zeta_k(\frac{1}{2}\delta_j^k-2\tilde\alpha\hat{{G}}^k_j)-\frac{1}{2}\frac{U}{r}(\frac{1}{2}\delta^i_j-2\tilde\alpha\hat{{G}}^i_j)\hat{\nabla}_i(\delta g_{tt}^{(0)}+\delta g_{rr}^{(0)})\,,
\end{align}
where we have used $\hat{\nabla}_i\hat{{G}}^i_j=0$ and $\tilde{\nabla}_i\tilde{{G}}^i_j=0$, which follow from the Bianchi identity for $h_{ij}^{(0)}$ and $\tilde{h}_{ij}^{(0)}\equiv h_{ij}^{(0)}+\delta g_{ij}$ respectively.

Defining $p\equiv -\frac{U}{r}\delta g_{rt}^{(0)}=-\frac{U}{2r}(\delta g_{rr}^{(0)}+\delta g_{tt}^{(0)})$ (and also using \eqref{incompress}) we obtain the final Stokes equation:
\begin{align}
&0=-\hat{\nabla}^k\hat{\nabla}_{(j}v_{k)}-(\frac{U}{r}\zeta_i-\hat{\nabla}_ip)(\frac{1}{2}\delta_j^i-2\tilde\alpha\hat{{G}}^i_j)\cr
&+\tilde\alpha\hat{\nabla}_i\Big[\frac{U}{r}\left(h^{ik}_{(0)}\hat{\nabla}_{(j}v_{k)}{h^l_l}^{(1)}+{h^i_j}^{(1)}\hat{\nabla}_kv^k-\hat{\nabla}^{(i}v^{k)}{h_{jk}}^{(1)}-{h^{ik}}^{(1)}\hat{\nabla}_{(j}v_{k)}+\delta^i_j(-{h_k^k}^{(1)}\hat{\nabla}_lv^l +{h^{kl}}^{(1)}\hat{\nabla}_kv_l)\right)\cr
&\qquad\qquad-4\hat{\nabla}_kv^k\hat{R}^i_j  +4\hat{\nabla}_{(j}v_{k)}\hat{{G}}^{ik}+4\hat{\nabla}^{(k}v^{i)}\hat{R}_{jk}+4\hat{\nabla}^{(k}v^{l)}{\,\hat{R}^i}_{kjl}\Big]\,.
\end{align}
Note that this depends only upon $v_i$, $p$, $\zeta_i$ and background quantities. 

We can also write this as an expansion in $\tilde\alpha$. As we derived in \eqref{eq:alpha_pert},
to leading order in $\tilde\alpha$ we have:
\begin{align}
h^{(1)}_{ij}=&\frac{2r}{U}\left( \hat{R}_{ij}-\frac{1}{D-2} V_0h^{(0)}_{ij}\right)\,,
\end{align}
where $V_0=-(D-1)(D-2)$. So this becomes
\begin{align}\label{stokeseqn}
0=&-\hat{\nabla}^k\hat{\nabla}_{(j}v_{k)}-(4\pi T\zeta_i-\hat{\nabla}_ip)(\frac{1}{2}\delta_j^i-2\tilde\alpha\hat{{G}}^i_j)\cr
&+\tilde\alpha\hat{\nabla}_i\Big[2(D-4) (D-1)h_{(0)}^{ik}\hat{\nabla}_{(j}v_{k)}+2\delta^i_j\hat{R}^{kl}\hat{\nabla}_kv_l+2\hat{\nabla}_{(j}v_{k)}\hat{R}^{ik}\nn
&\qquad\qquad+2\hat{\nabla}^{(k}v^{i)}\hat{R}_{jk}+4\hat{\nabla}^{(k}v^{l)}{\,\hat{R}^i}_{kjl}\Big]+\mathcal{O}(\tilde\alpha^2)\,.
\end{align}

\subsubsection{Hamiltonian Constraint}
The Hamiltonian constraint, $H=0$, where $H$ is given in \eqref{hcon},
can be evaluated at the horizon
and we find
\begin{align}
H
=&{\sqrt{h_{(0)}}}\frac{U^{1/2}}{r}(\delta_i^j-4\tilde\alpha \hat{{G}}_i^j)\hat{\nabla}_{j}v^{i}+\mathcal{O}(r^{1/2})\,.
\end{align}
Thus, the Hamiltonian constraint leads to the same condition given in \eqref{incompress}.


\providecommand{\href}[2]{#2}\begingroup\raggedright\endgroup

\end{document}